*Special Coverings of Sets and Boolean Functions*

Stepan G. Margaryan

e-mail: Stepan-Margar@mskh.am
stmargaryan@gmail.com

We will study some important properties of Boolean functions based on newly introduced concepts called ''Special Decomposition of a Set'' and ''Special Covering of a Set''.

An important result of these concepts is the equivalence of the Boolean satisfiability problem and the problem of the existence of a special cover for a set.

The introduced concepts easily solve the question of how change in clauses affect the satisfiability of a function. They easily determine from which clause and which literal can be removed or added to the clause to preserve the satisfiability of the function.

The concept of generating a satisfiable function by another satisfiable function through admissible changes in the function's clauses is also introduced. If the generation of a function by another function is defined as a binary relation, then the set of satisfiable functions of $n$ variables, represented in conjunctive normal form with $m$ clauses is partitioned to equivalence classes.

Moreover, we prove that any two satisfiable Boolean functions of $n$ variables, represented in conjunctive normal form with $m$ clauses, can be generated from each other in polynomial time.

## Introduction

Despite the fact that the field of Boolean functions has long been widely and well-studied, we obtain important and interesting results in this field based on newly introduced concepts of special decomposition and special covering of a set.

These concepts are flexible and simple in use, so they enable us to study important problems concerning Boolean functions represented in conjunctive normal form, including the satisfiability problem. We will explore the possibility of covering a set with specially chosen its subsets, as well as the application of these concepts to Boolean functions.

We will show that any Boolean function represented in conjunctive normal form generates a special decomposition of the set of clauses of this function, and any special decomposition of the set generates a Boolean function in conjunctive normal form.

Moreover, we prove that these generations run in polynomial time.

In addition, we prove that the Boolean function in conjunctive normal form is satisfiable if and only if there is a special covering for the set of clauses of this function under the corresponding decomposition. So, these problems are polynomially equivalent.

One of the main goals of this article is to explore the possibilities of the concepts of special decomposition and special covering, using them in relation to satisfiable functions.

We introduce and study the concept of admissible changes in the subsets included in special decomposition of a set. We show that any admissible change in these subsets will carry to admissible changes in the clauses of the function generated by this special decomposition. Note that changes in clauses will mean adding a literal to a clause, removing a literal from a clause, or swapping negative and positive literals of the same variable.

Typically, when transforming a Boolean function given in conjunctive normal form, it is not always obvious what impact a change in a clause may have on the satisfiability of the function. Furthermore, it is not always easy to determine from which clause and which literal can be removed or added to the clause, in order to preserve the satisfiability of the function.

This problem for a function is easily solved if we use the special decomposition generated by this function. The definition of admissible changes allows us to transform some clauses of a satisfiable function so that the resulting function remains satisfiable.

To study the results, we introduce the concept of generation of a satisfiable function by another satisfiable function. Using this concept, as a result of any step of admissible changes, new satisfiable function is obtained. We also prove that the procedure for generation of a function by another function has a polynomial time complexity.

Further, defining the generation of a function by another function as a binary relation, we prove that the set of satisfiable functions of $n$ variables represented in conjunctive normal form with $m$ clauses, is partitioned into equivalence classes. All functions included in the same class have a common satisfiable assigning tuple.

Moreover, extending the rules of admissible changes, we prove an important result:

If two arbitrary Boolean functions of $n$ variables represented in conjunctive normal form with $m$ clauses, are both satisfiable, then either can be generated by the other using extended admissible changes in polynomial time.

In other words, we prove the following interesting result.

If we are given a satisfiable Boolean function, then as a result of sequential application of the operations of the extended admissible changes, the following will be true:

1) we get a chain of satisfiable functions,

2) such a chain exists between an arbitrary pair of satisfiable functions. The generation procedure starts with one of these functions and ends with another function.

3) given an arbitrary pair of satisfiable functions, the generation procedure any of the functions by another function is performed in polynomial time.

## 1. *Basic Concepts*

We will deal with different nonempty sets. These sets are assumed to be ordered unless otherwise stated. Let the set $S = \{e_1, e_2, \ldots, e_m\}$ be given.

It is assumed that for some natural number $n$, $n$ arbitrary ordered pairs of arbitrary subsets of the set $S$ are given. We denote these ordered pairs by
$$(M_1^0, M_1^1), (M_2^0, M_2^1), \ldots, (M_n^0, M_n^1).$$

It is important to note that the pairs are numbered in no particular order.

We will use the notation $d_n S$ for an arbitrarily ordered set of these ordered pairs:
$$d_n S = \{(M_1^0, M_1^1), (M_2^0, M_2^1), \ldots, (M_n^0, M_n^1)\}.$$
The Boolean values $\alpha_i$ and $\bar{\alpha}_i$, for $i \in \{1, \ldots, n\}$, will be used to denote superscripts of subsets:
$$\bar{\alpha}_i = 0 \text{ if } \alpha_i = 1, \text{ and } \bar{\alpha}_i = 1 \text{ if } \alpha_i = 0.$$
The superscript tuple $(\alpha_1, \alpha_2, \ldots, \alpha_n)$ of subsets will also be called a Boolean tuple.

*Definition* 1.1. The set $d_n S$ will be called a special decomposition of the set $S$, if
(1.1.1) $\quad \forall i \in \{1, \ldots, n\} \ (M_i^0 \cap M_i^1) = \emptyset$,
(1.1.2) $\quad \forall i \in \{1, \ldots, n\} \ (M_i^0 \neq \emptyset \text{ or } M_i^1 \neq \emptyset)$,
(1.1.3) $\quad \bigcup_{i=1}^n (M_i^0 \cup M_i^1) = S$.

Obviously, the same subsets of the set can form different special decompositions and also these subsets may not allow any special decompositions.

*Definition* 1.2. Let the set $d_n S$ be a special decomposition of the set $S$.
For any Boolean tuple $(\alpha_1, \alpha_2, \ldots, \alpha_n)$ the ordered set
$$c_n S = \{M_1^{\alpha_1}, M_2^{\alpha_2}, \ldots, M_n^{\alpha_n}\}$$
will be called a special covering for the set $S$ under the decomposition $d_n S$, if
$$\bigcup_{i=1}^n M_i^{\alpha_i} = S.$$

It follows from the Definition 1.2, that for any $i \in \{1, \ldots, n\}$ the subsets $M_i^0$ and $M_i^1$ cannot simultaneously belong to the covering, but one of them exactly belongs.

*Proposition* 1.3. Let for some Boolean tuple $(\alpha_1, \alpha_2, \ldots, \alpha_n)$, the set
$$c_n S = \{M_1^{\alpha_1}, M_2^{\alpha_2}, \ldots, M_n^{\alpha_n}\}$$
be a special covering for the set $S$ under the special decomposition $d_n S$.
If $M_i^\alpha \not\subseteq \bigcup_{j \neq i}(M_j^0 \cup M_j^1)$ for some $\alpha \in \{0,1\}$, then $M_i^\alpha \in c_n S$.

*Proof.* Suppose that $M_i^\alpha \notin c_n S$. It means that $M_i^{\bar{\alpha}} \in c_n S$. Since, by condition
$$M_i^\alpha \not\subseteq \bigcup_{j \neq i}(M_j^0 \cup M_j^1),$$
then there exists an element $e \in M_i^\alpha$, such that for any subsets $M_j^{\alpha_j}$ and $M_j^{\bar{\alpha}_j}$ if $j \neq i$ then
$$(e \notin M_j^{\alpha_j}) \ \& \ (e \notin M_j^{\bar{\alpha}_j}).$$
On the other hand, since $M_i^\alpha \cap M_i^{\bar{\alpha}} = \emptyset$, then it follows from $e \in M_i^\alpha$ that $e \notin M_i^{\bar{\alpha}}$.
So, $c_n S$ cannot be a special covering for the set $S$, since no subset included in it contains the element $e$. And this is a contradiction. $\nabla$
(By the symbol $\nabla$ we mark the end of the proof).

*Corollary* 1.3.1. If under some special decomposition $d_n S$ of the set $S$ there is an ordered pair $(M_i^0, M_i^1) \in d_n S$ such, that
$$M_i^0 \not\subseteq \bigcup_{j \neq i}(M_j^{\alpha_j} \cup M_j^{\bar{\alpha}_j}) \text{ and } M_i^1 \not\subseteq \bigcup_{j \neq i}(M_j^{\alpha_j} \cup M_j^{\bar{\alpha}_j}),$$
then there is no special covering for the set $S$ under the decomposition $d_n S$.

*Proof.* If, under the given conditions, there were a special covering, then, according to Proposition 1.3, it would have to include both subsets $M_i^0$ and $M_i^1$. But this is contrary to the Definition 1.2. $\nabla$

Let $d_nS = \{(M_1^0, M_1^1), \ldots, (M_n^0, M_n^1)\}$ be an ordered set of arbitrary ordered pairs of subsets of the set $S$ (for some natural number $n$).

For any $1 \leq k \leq n$, the permutation of the components of ordered pairs
$$(M_{i_1}^0, M_{i_1}^1), \ldots, (M_{i_к}^0, M_{i_к}^1)$$
of the set $d_nS$, when the orders of the elements of $d_nS$ do not change, will be called an $I$-transformation of the set $d_nS$.

The ordered set obtained as a result of $I$-transformation will be denoted as
$$(i_1, i_2, \ldots, i_k)I(d_nS).$$

If it is not necessary to mark the numbers of the pairs participating in the transformation, then $I$-transformation of the set $d_nS$ will be denoted by $I(d_nS)$:
$$I(d_nS) = \{(M_1^{\alpha_1}, M_1^{\bar{\alpha}_1}), \ldots, (M_i^{\alpha_i}, M_i^{\bar{\alpha}_i}), \ldots, (M_n^{\alpha_n}, M_n^{\bar{\alpha}_n})\},$$
for a Boolean tuple $(\alpha_1, \ldots, \alpha_n)$.

The ordered pairs of this decomposition are defined as follows:
$(M_i^{\alpha_i}, M_i^{\bar{\alpha}_i}) = (M_i^0, M_i^1)$, if the components of the $i$-th pair are not displaced,
$(M_i^{\alpha_i}, M_i^{\bar{\alpha}_i}) = (M_i^1, M_i^0)$, if the components of the $i$-th pair are displaced.

*Lemma* 1.4. If $d_nS$ is an ordered set of ordered pairs of subsets of the set $S$ then for any $I$-transformation $I(d_nS)$, the following is true:

i) $d_nS$ is a special decomposition of the set $S$ if and only if $I(d_nS)$ is a special decomposition of the set $S$.

(ii) If $d_nS$ is a special decomposition of the set $S$, then there exists a special covering for the set $S$ under the decomposition $d_nS$ if and only if it exists under the decomposition $I(d_nS)$.

*Proof*. i) During the transition from the set $d_nS$ to the set $I(d_nS)$ and the transition from $I(d_nS)$ to $d_nS$, the contents of the subsets of decomposition do not change. Only the orders of the components of some ordered pairs change.

Therefore, the sets $d_nS$ and $I(d_nS)$ are either at the same time special decompositions of the set $S$, or at the same time they are not such decompositions. Also, it is obvious that if under decomposition $d_nS$, the set $c_nS = \{M_1^{\alpha_1}, \ldots, M_n^{\alpha_n}\}$ ($\alpha_i \in \{0, 1\}$) is a special covering for the set $S$, then it will also be a special covering for the set $S$ under decomposition $I(d_nS)$ and vice versa. So, the points i) and ii) are true. ∇

According to Lemma 1.4, for any special decomposition of the set $S$, any $I$-transformation preserve the possibility of being a special decomposition of the set $S$ and having a special covering for $S$ under such a decomposition.

We will distinguish the subsets included in ordered pairs according to the order of their location in these pairs.

Let for Boolean tuple $(\alpha_1, \ldots, \alpha_n)$, $d_nS$ be a special decomposition:
$$d_nS = \{(M_1^{\alpha_1}, M_1^{\bar{\alpha}_1}), \ldots, (M_n^{\alpha_n}, M_n^{\bar{\alpha}_n})\}.$$
The subsets $M_1^{\alpha_1}, \ldots, M_n^{\alpha}$ will be called the subsets of 0-domain,
The subsets $M_1^{\bar{\alpha}_1}, \ldots, M_n^{\bar{\alpha}_n}$ will be called the subsets of 1-domain.

If the components of an ordered pair $(M_i^0, M_i^1)$ are permuted, then the subset $M_i^1$ becomes a subset of the 0-domain, and the subset $M_i^0$ becomes a subset of the 1-domain.

Thus, for technical convenience, for any $\alpha \in \{0, 1\}$ we denote:

$M^\alpha = \bigcup_{i=1}^n M_i^\alpha$,

$sM^\alpha = \{M_1^\alpha, M_2^\alpha, \ldots, M_n^\alpha\}$,

$(i_1, \ldots, i_k)sM^\alpha$ is the set obtained by replacing the subsets $M_{i_1}^\alpha, \ldots, M_{i_k}^\alpha$ with the subsets $M_{i_1}^{\bar\alpha}, \ldots, M_{i_k}^{\bar\alpha}$, respectively, in the set $sM^\alpha$.

Note that $sM^\alpha$ and $(i_1, \ldots, i_k)sM^\alpha$ we consider as ordered sets.

*Definition* 1.5. For any $\alpha \in \{0, 1\}$:

i) the set $sM^\alpha$ will be called a set of $\alpha$-components of ordered pairs of the decomposition.

ii) For any $\{i_1, \ldots, i_k\} \subseteq \{1, \ldots, n\}$ the set $(i_1, \ldots, i_k)sM^\alpha$ is called a set of $\alpha$-components of the ordered pairs of decomposition $(i_1, \ldots, i_k)I(d_n S)$.

iii) For any decomposition, the set of $\alpha$-components of ordered pairs will also be called a set of subsets of the $\alpha$-domain.

iv) If the set of subsets of the $\alpha$-domain is a special covering for the set $S$, then such a covering will be called a special $M^\alpha$-covering or briefly $M^\alpha$-covering for the set $S$.

*Lemma* 1.6. Let the set $d_n S = (M_1^0, M_1^1), \ldots, (M_i^0, M_i^1), \ldots, (M_n^0, M_n^1)\}$ be a special decomposition of the set $S$.

Then, there exists a special covering for the set $S$ under the special decomposition $d_n S$ if and only if for some $\alpha \in \{0,1\}$ there exists an $M^\alpha$-covering for the set $S$ under some special decomposition $I(d_n S)$.

*Proof.* Obviously, for any $\alpha \in \{0,1\}$, the procedure for forming the $\alpha$-domain does not violate the Definition 1.2 of a special covering. Therefore, an $M^\alpha$-covering is also a special covering for the set $S$.

Now suppose that there is a special covering for the set $S$. Let it be the set
$$c_n S = \{M_1^{\alpha_1}, M_2^{\alpha_2}, \ldots, M_n^{\alpha_n}\}.$$

If $M_i^{\alpha_i} \in sM^\alpha$ for any $i \in \{1, \ldots, n\}$, then $c_n S$ is also an $M^\alpha$-covering.

If $M_{j_1}^{\bar\alpha}, \ldots, M_{j_l}^{\bar\alpha}$ are subsets such, that $(\{M_{j_1}^{\bar\alpha}, \ldots, M_{j_l}^{\bar\alpha}\} \subseteq sM^{\bar\alpha})$ & $(\{M_{j_1}^{\bar\alpha}, \ldots, M_{j_l}^{\bar\alpha}\} \subseteq c_n S)$, then applying $I$-transformation with respect to the ordered pairs $(M_{j_1}^\alpha, M_{j_1}^{\bar\alpha}), \ldots, (M_{j_l}^\alpha, M_{j_l}^{\bar\alpha})$, according to Lemma 1.4, we obtain that $c_n S$ is also an $M^\alpha$-covering for the set $S$. $\nabla$

## 2. *Boolean Functions and Special Decompositions*

Let for natural numbers $n$ and $m$, $f(x_1, x_2, \ldots, x_n)$ be a Boolean function of $n$ variables represented in conjunctive normal form ($CNF$) with $m$ clauses.

We assume that the clauses of the function are numbered in some natural manner.

Let $c_i$ be some number corresponding to the $i$-th clause of the function in some natural one-to-one correspondence. We will identify the clause with its number if it does not lead to ambiguity.

Thus, for some $k \in \{1, \ldots, n\}$ and $\{j_1, \ldots, j_k\} \subseteq \{1, \ldots, n\}$, we will use the notation
$$c_i = x_{j_1}^{\alpha_1} \vee \ldots \vee x_{j_k}^{\alpha_k}, \text{ where } \alpha_j \in \{0,1\}, \ x_j^0 = \bar x_j, \ x_j^1 = x_j, \ j \in \{1, \ldots, n\}.$$

With this notation, the function $f(x_1, x_2, \ldots, x_n)$ will be represented as
$$f(x_1, x_2, \ldots, x_n) = \bigwedge_{i=1}^m c_i.$$

For simplicity and technical convenience, we assume the following:
- no variable and its negation are included in any clause simultaneously,
- if the function contains $n$ variables, then they are numbered sequentially. That is, for any $j \in \{1, \ldots, n\}$, the literal $x_j^\alpha$ appears in some clauses for some $\alpha \in \{0,1\}$.

Obviously, this assumption does not limit the set of functions being considered.

*We say that the clauses of the set $\{c_{j_1}, \ldots, c_{j_k}\}$ are satisfiable if there is a Boolean assignment tuple $(\sigma_1, \ldots, \sigma_n)$, such that any of these clauses takes the value $1$ when the variables $x_1, \ldots, x_n$ are assigned the values $\sigma_1, \ldots, \sigma_n$, respectively.*

2.1. *Special Decomposition of Clauses of a Boolean Function.*

Let $S(f) = \{c_1, c_2, \ldots, c_m\}$ be the set of clauses of the function $f(x_1, \ldots, x_n)$.

Further we will consider $S(f)$ as an ordered set. It is easy to see that this does not prevent us from considering any Boolean function in conjunctive normal form.

Based on the clauses of the function $f(x_1, \ldots, x_n)$, we form the subsets of the set $S(f)$.

For any $i \in \{1, \ldots, n\}$ and $\alpha \in \{0,1\}$ we denote by $F_i^\alpha$ and $F_i^{\bar\alpha}$ the subsets of the set $S(f)$.

$F_i^0 = \{c_j \,/\, c_j \in S(f) \text{ and } c_j \text{ contains the literal } \bar{x}_i, \ (j \in \{1, \ldots, m\})\}$.

$F_i^1 = \{c_j \,/\, c_j \in S(f) \text{ and } c_j \text{ contains the literal } x_n, \ (j \in \{1, \ldots, m\})\}$.

Subsets of clauses of a function will be denoted by capital letters corresponding to the function designation. Let's form the following ordered set of ordered pairs of these subsets:

$$d_n S(f) = \{(F_1^0, F_1^1), (F_2^0, F_2^1), \ldots, (F_n^0, F_n^1)\}.$$

*We will say that the ordered set $d_n S(f)$ is a decomposition of the set $S(f)$ generated by the conjunctive normal form of the function $f(x_1, \ldots, x_n)$.*

*Lemma* 2.2. For any function $f(x_1, \ldots, x_n)$, represented in conjunctive normal form, the set $d_n S(f)$ is a special decomposition of the set $S(f)$.

*Proof.* Consider the conditions (1.1.1), (1.1.2) and (1.1.3).

(1.1.1) $\forall i \in \{1, \ldots, n\}(F_i^0 \cap F_i^1) = \emptyset$.

This is evident since none of the clauses contains the literals $\bar{x}_i$ and $x_i$ simultaneously.

(1.1.2) $\forall i \in \{1, \ldots, n\} \ F_i^0 \neq \emptyset$ or $F_i^1 \neq \emptyset$)

If $F_i^0 = \emptyset$ and $F_i^1 = \emptyset$ for some $i \in \{1, \ldots, n\}$, then the literals $\bar{x}_i$ and $x_i$ do not belong to any clause. And this contradicts the formation of the subsets $F_i^0$ and $F_i^1$.

(1.1.3) $\bigcup_{i=1}^{n}(F_i^0 \cup F_i^1) = S(f)$,

Let for some $j \in \{1, \ldots, m\}$, $c_j \in \bigcup_{i=1}^{n}(F_i^0 \cup F_i^1)$.

Then, for some $i \in \{1, \ldots, n\}$, $c_j \in F_i^0$ or $c_j \in F_i^1$, which means that $c_j \in S(f)$.

Let for some $j \in \{1, \ldots, m\}$, $c_j \in S(f)$.

Since $c_j$ is not an empty clause, then it contains some literals. So, for some $i \in \{1, \ldots, n\}$ either $\bar{x}_i$ is included in the clause $c_j$, or $x_i$ is included in the clause $c_j$. But then $c_j \in F_i^0$ or $c_j \in F_i^1$, which means that $c_j \in \bigcup_{i=1}^{n}(F_i^\alpha \cup F_i^{\bar\alpha})$.

Therefore, for any function $f(x_1, \ldots, x_n)$, represented as conjunctive normal form, the set $d_n S(f)$ is a special decomposition of the set $S(f)$. ∇

If under the special decomposition $d_n S(f)$, there exists a special covering for the set $S(f)$, then we will denote such a covering by
$$c_n S(f) = \{F_1^{\alpha_1}, F_2^{\alpha_2}, \ldots, F_n^{\alpha_n}\}.$$

*Theorem* 2.3. For any Boolean function $f(x_1, \ldots, x_n)$ represented in conjunctive normal form, the following is true:

There is a Boolean assigning tuple $(\sigma_1, \ldots, \sigma_n)$ such that $f(\sigma_1, \ldots, \sigma_n) = 1$ if and only if there is a special covering for the set $S(f)$ under the decomposition $d_n S(f)$.

*Proof.* Let $f(\sigma_1, \ldots, \sigma_n) = 1$ for some assigning tuple $(\sigma_1, \ldots, \sigma_n)$.

We will show that then the set $c_n S(f) = \{F_1^{\sigma_1}, F_2^{\sigma_2}, \ldots, F_n^{\sigma_n}\}$ will be a special covering for the set $S(f)$ under the special decomposition $d_n S(f)$.

To show this, we prove that $\bigcup_{i=1}^{n} F_i^{\sigma_i} = S(f)$.

It is enough to show that each clause belongs to some subset included in the set $c_n S(f)$.

Suppose that there is a clause $c_j \in S(f)$ that does not belong to any of the subset included in $c_n S(f)$. It means that none of the literals $x_1^{\sigma_1}, x_2^{\sigma_2}, \ldots, x_n^{\sigma_n}$ is included in the clause $c_j$.

Therefore, $c_j$ is the disjunction of some literals of the form $x_i^{\bar{\sigma}_i}$.

Since $\sigma_i^{\bar{\sigma}_i} = 0$ for any $i \in \{1, \ldots, n\}$, then for given values of variables, the clause $c_j$ will take the value 0. This contradicts the assumption that $f(\sigma_1, \sigma_2, \ldots, \sigma_n) = 1$.

So, each clause is included in some subset included in the set $c_n S(f)$.

Let for some superscript tuple $(\alpha_1, \alpha_2, \ldots, \alpha_n) \in \{0,1\}$ the set
$$c_n S(f) = \{F_1^{\alpha_1}, F_2^{\alpha_2}, \ldots, F_n^{\alpha_n}\}$$
is a special covering for the set $S(f)$ under the decomposition $d_n S(f)$.

By definition, the subset $F_i^{\alpha_i}$ contains clauses that contain the literal $x_i^{\alpha_i}$. It is easy to see that if $x_i^{\alpha_i} = 1$, then the value of all clauses included in the set $F_i^{\alpha_i}$ will be equal to 1,

That is, for any $i \in \{1, \ldots, n\}$ and $j \in \{1, \ldots, m\}$, if $(x_i^{\alpha_i} = 1)$ & $(c_j \in F_i^{\alpha_i})$ then $(c_j = 1)$.

Obviously, if $\sigma_1 = \alpha_1, \sigma_2 = \alpha_2, \ldots, \sigma_n = \alpha_n$, then $f(\sigma_1, \ldots, \sigma_n) = 1$. $\nabla$

2.4. *Generation of the Boolean Function Based on a Special Decomposition.*

Let's now form a Boolean function, represented in conjunctive normal form, based on some special decomposition $d_n S = \{(M_1^0, M_1^1), \ldots, (M_i^0, M_i^1), \ldots, (M_n^0, M_n^1)\}$.

We denote this function as $h(x_1, \ldots, x_n)$, where $x_1, \ldots, x_n$ are Boolean variables.

To form the function $h(x_1, \ldots, x_n)$, first, for any element $e_i \in S$, we form the set of literals, denoted by $l(e_i)$, based on the positions of the subsets containing the element $e_i$.

That is, for any $j \in \{1, \ldots, n\}$ and $\alpha \in \{0,1\}$, if $e_i \in M_j^\alpha$, then we form the literal $x_j^\alpha$ and add it to the formed set $l(e_i)$.

It is easy to see, that when forming the literals $x_j^\alpha$, the number of variables will be $n$.

In fact, for each element $e_i \in \{e_1, e_2, \ldots, e_m\}$ we will have:
$$l(e_i) = \{x_j^\alpha \,/\, e_i \in M_j^\alpha, \ j \in \{1, \ldots, n\}, \ \alpha \in \{0,1\}\}.$$

Let $c_i$ be the clause formed by the literals of the set $l(e_i)$. Obviously, the number of these clauses will be equal to $m$. Then, we form the function $h$ as follows:
$$h(x_1, \ldots, x_n) = \bigwedge_{i=1}^{m} c_i.$$

We say that the function $h(x_1, \ldots, x_n)$ is generated by the special decomposition $d_n S$.

It is easy to see that the generated function $h(x_1, \ldots, x_n)$ is a Boolean function in conjunctive normal form. It is also obvious that particular function in conjunctive normal form will correspond to any special decomposition.

*Theorem* 2.5. If the set $d_n S = \{(M_1^0, M_1^1), \ldots, (M_n^0, M_n^1)\}$ be a special decomposition of the set $S$, and $h(x_1, \ldots, x_n)$ is the function generated by this decomposition, then:

There exists a special covering for the set $S$ under the decomposition $d_n S$, if and only if there exists a Boolean assignment tuple $(\sigma_1, \ldots, \sigma_n)$ such that $h(\sigma_1, \ldots, \sigma_n) = 1$.

*Proof.* Suppose that for some superscript tuple $(\alpha_1, \alpha_2, \ldots, \alpha_n)$, the set
$$c_n S = \{M_1^{\alpha_1}, M_2^{\alpha_1}, \ldots, M_n^{\alpha_n}\}$$
is the special covering for the set $S$.

This means that for any $e_i \in S$, there exists a subset $M_j^{\alpha_j} \in c_n S$ such that $e_i \in M_j^{\alpha_j}$. But then, by definition, the literal $x_j^{\alpha_j}$ is included in the clause $c_i$. That is, for any $i \in \{1, \ldots, m\}$, if $e_i \in M_j^{\alpha_j}$ then the literal $x_j^{\alpha_j}$ is included in the clause $c_i$.

It is easy to notice, that if $\sigma_1 = \alpha_1, \ldots, \sigma_n = \alpha_n$, then $h(\sigma_1, \ldots, \sigma_n) = 1$.

Suppose now, there is a Boolean assigning tuple $(\sigma_1, \ldots, \sigma_n)$ such that $h(\sigma_1, \ldots, \sigma_n) = 1$.

According to Theorem 2.3, the set $c_n S(h) = \{H_1^{\sigma_1}, H_2^{\sigma_2}, \ldots, H_n^{\sigma_n}\}$ is a special covering for the set $S(h)$ under the decomposition $d_n S(h)$.

Let us prove that then the set $c_n S = \{M_1^{\sigma_1}, M_2^{\sigma_2}, \ldots, M_n^{\sigma_n}\}$ will be a special covering for the set $S$. Since the set $c_n S(h)$ is a special covering for the set $S(h)$, for any clause $c_i$ there exists a subset $H_j^{\sigma_j} \in c_n S(h)$ such that $c_i \in H_j^{\sigma_j}$. This means that the clause $c_i$ contains the literal $x_j^{\sigma_j}$, since by definition
$$H_j^{\sigma_j} = \{c_k \,/\, c_k \in S(h) \text{ and } c_k \text{ contains } x_j^{\sigma_j}, (k \in \{1, \ldots, m\})\}.$$

On the other hand, by definition the clause $c_i$ contains the literal $x_j^{\sigma_j}$ only if $e_i \in M_j^{\sigma_j}$.

Since each element $e_i \in S$ determines the composition of one clause, and each clause is defined by one element of the set $S$, then it is easy to prove that for any element $e_i \in S$ there exists a subset $M_j^{\sigma_j} \in c_n S$ such, that $e_i \in M_j^{\sigma_j}$.

Therefore, the set $c_n S = \{M_1^{\sigma_1}, M_2^{\sigma_2}, \ldots, M_n^{\sigma_n}\}$ is a special covering for the set $S$. $\nabla$

In fact, we have established an important relationship between the Boolean satisfiability problem and the problem of finding a special covering for a set.

-each Boolean function $f(x_1, \ldots, x_n)$ of $n$ variables represented in conjunctive normal form with $m$ clauses, generates a special decomposition $d_n S(f)$ of the set $S(f)$ of $m$ elements.

- each special decomposition of any set of $m$ elements and containing $n$ ordered pairs, generates a Boolean function of $n$ variables in conjunctive normal form with $m$ clauses.

Using the Theorems 2.3 and 2.5, this means that any decidability result for any of these problems leads to the same result for other.

Later we will estimate the number of operations required to perform any of these procedures.

We will show that these procedures have polynomial time complexity with respect to the corresponding input data.

## 3. *Some Important Properties of Special Decompositions.*

To study some of important properties of the concepts of special decomposition and special covering, we introduce some transformations in the special decomposition by changing the contents of subsets such that the conditions of the special decomposition are preserved. The transformations will be made based on some special covering for the given set under the given special decomposition. This will mean that during the changes, the contents of the subsets included in the original special covering may also change, but their superscripts and subscripts are preserved

We assume that the ordered set
$$d_n S = \{(M_1^0, M_1^1), \ldots, (M_i^0, M_i^1), \ldots, (M_n^0, M_n^1)\}$$
is a special decomposition for a set $S$, and for a Boolean tuple $(\alpha_1, \ldots, \alpha_n)$, the ordered set
$$c_n S = \{M_1^{\alpha_1}, \ldots, M_i^{\alpha_i}, \ldots, M_n^{\alpha_n}\}$$
is some special covering for the set $S$ under the special decomposition $d_n S$.

We will use the notation $M_i^\delta \in (M_i^0, M_i^1)$, meaning that $M_i^\delta = M_i^0$ or $M_i^\delta = M_i^1$.

<u>*Definition*</u> 3.1. (i) Let the ordered pairs $(M_i^\alpha, M_i^{1-\alpha})$ and $(M_j^\beta, M_j^{1-\beta})$ are included in the special decomposition $d_n S$, $M_i^\delta \in (M_i^0, M_i^1)$ and $M_j^\gamma \in (M_j^0, M_j^1)$, for some $i, j \in \{1, \ldots, n\}$.

We say that the changes in the contents of the subsets $M_j^\gamma$ and $M_i^\delta$ are admissible under the tuple $(\alpha_1, \ldots, \alpha_n)$, if the changes are made in accordance with the following points:

(i.1) for an element $e \in M_i^\delta$, the subset $M_i^\delta$ is replaced with the set $M_i^\delta \setminus \{e\}$ in the ordered pair $(M_i^0, M_i^1)$, if $M_i^\delta \notin c_n S$ and $((M_i^\delta \setminus \{e\}) \cup M_i^{\bar{\delta}}) \neq \emptyset$.

(i.2) for an element $e \notin (M_i^0 \cup M_i^1)$, the subset $M_i^\delta$ is replaced with the set $M_i^\delta \cup \{e\}$ in the ordered pair $(M_i^0, M_i^1)$.

(i.3) if the subsets $M_j^\gamma$ and $M_i^\delta$ are both included in $c_n S$, then for an element $e$ such that $e \in M_j^\gamma$ and $e \notin M_i^{\bar{\delta}}$, the subset $M_j^\gamma$ is replaced with the set $M_j^\gamma \setminus \{e\}$ and the subset $M_i^\delta$ is replaced with the set $M_i^\delta \cup \{e\}$ in the corresponding ordered pairs, respectively.

(ii) We say that the ordered set $d_n SG$ is generated by the decomposition $d_n S$ as a result of admissible changes under the tuple $(\alpha_1, \ldots, \alpha_n)$, if these changes are made in the components of some ordered pairs included in the decomposition $d_n S$, in accordance with points (i.1) - (i.3).

(iii) We say that the ordered set $c_n SG$ is generated as a result of admissible changes under the tuple $(\alpha_1, \ldots, \alpha_n)$ in the special decomposition $d_n S$, if $c_n SG$ is a set that matches the ordered set corresponding to $c_n S$ in the resulting decomposition.

<u>*Remark*</u> 3.2. About to the point (i.3) of this definition:

For the subsets $M_j^\gamma$, $M_i^\delta \in c_n S$, an element $e$ moves from the subset $M_j^\gamma$ to the subset $M_i^\delta$ provided that $e \notin M_i^{\bar{\delta}}$. It is easy to see, that in case of $e \in M_i^\delta$, we actually obtain the removal of the element $e$ from the subset $M_j^\gamma$.

That is, this point gives us the opportunity, if necessary, to remove an element from a subset included in the special covering.

Sometimes, if it does not lead to ambiguity, we will use the notation $c_n SG$ for the ordered set which either coincides with the set $c_n S$ or is generated by the sets $d_n S$ and $c_n S$.

*Theorem* 3.3. Let for some Boolean tuple $(\alpha_1, \ldots, \alpha_n)$, the ordered set
$$c_n S = \{M_1^{\alpha_1}, \ldots, M_i^{\alpha_i}, \ldots, M_n^{\alpha_n}\}$$
be a special covering for the set $S$ under the special decomposition $d_n S$.

If the ordered sets $d_n SG$ and $c_n SG$ are generated as a result of admissible changes under the tuple $(\alpha_1, \ldots, \alpha_n)$ in the decomposition $d_n S$, then:
- $d_n SG$ is also a special decomposition of the set $S$,
- $c_n SG$ is a special covering for the set $S$ under the decomposition $d_n SG$.

*Proof.* It is easy to see that the admissible changes in the special decomposition $d_n S$ do not violate the conditions of Definition 1.1 of the special decomposition.

So, $d_n SG$ is a special decomposition of the set $S$.

Consider the ordered set $c_n SG$. Obviously, this is an ordered set with the same numbering and with same superscripts of elements as in the ordered set $c_n S$.

Suppose that $c_n SG$ coincides with $c_n S$. This means that none of the subsets included in $c_n S$ has changed during the admissible changes. Other changes do not affect the special covering, so $c_n SG$ is a special covering for the set $S$ under the decomposition $d_n SG$.

Let the ordered set $c_n SG$ be generated by the special covering $c_n S$.

Obviously, removing an element from some subset that is not included in the set $c_n S$ cannot affect the special covering.

On the other hand, according to Definition 3.1(i.3), we remove an element from some subset included in the set $c_n S$ only if this element is added to another subset included in $c_n S$.

Therefore, the contents of the subsets included in $c_n S$ may change, but in general the elements included in the subsets included in $c_n SG$ will be the same as those in the set $c_n S$.

In addition, $(\alpha_1, \ldots, \alpha_n)$ will be the tuple of the superscripts also for the subsets included in the ordered set $c_n SG$.

Thus, the set $c_n SG$ covers the set $S$ under the special decomposition $d_n SG$. $\nabla$

## 4. *Admissible Changes in Clauses of Functions.*

Consider a Boolean function $f(x_1, \ldots, x_n)$ of $n$ variables represented in conjunctive normal form with $m$ clauses and recall some important results.

Lemma 2.2 states that any Boolean function represented in $CNF$ generates a special decomposition of the set $S(f)$:
$$d_n S(f) = \{(F_1^0, F_1^1), (F_2^0, F_2^1), \ldots, (F_n^0, F_n^1)\}.$$

According to Theorem 2.3, the function $f(x_1, \ldots, x_n)$ is satisfiable if and only if there is a special covering for the set $S(f)$ under the special decomposition $d_n S(f)$.

Actually, it is proven that there is a Boolean tuple $(\sigma_1, \ldots, \sigma_n)$ such that $f(\sigma_1, \ldots, \sigma_n) = 1$ if and only if the ordered set
$$c_n S(f) = \{F_1^{\sigma_1}, F_2^{\sigma_2}, \ldots, F_n^{\sigma_n}\},$$
is a special covering for the set $S(f)$ under the special decomposition $d_n S(f)$. That is, the satisfiable assignment tuple defines the subsets that cover the set $S(f)$, and vice versa.

Also, section 2.4 describes a procedure that, based on any special decomposition, generates a Boolean function in conjunctive normal form.

Similar to the definition (3.1(ii)), we will use the notation $d_n S(f)G$ for the decomposition generated as a result of admissible changes under the tuple $(\sigma_1, \ldots, \sigma_n)$ in the decomposition $d_n S(f)$.

Also, similar to the Definition 3.1(iii) we will use the notation $c_n SG(f)$ for the ordered set generated as a result of admissible changes under the special covering $c_n S(f)$ in the special decomposition $d_n S(f)$.

In fact, the ordered set $c_n SG(f)$ either coincides with the ordered set $c_n S(f)$ or is obtained by applying the points (i.2) and (i.3) of the Definition 3.1.

*Definition* 4.1. Let the Boolean function $f(x_1, \ldots, x_n)$ of $n$ variables be represented in conjunctive normal form and let it be a satisfiable function.

We say that the function $h(x_1, \ldots, x_n)$ is generated by the function $f(x_1, \ldots, x_n)$ as a result of admissible changes under the assignment tuple $(\sigma_1, \ldots, \sigma_n)$, if the following conditions are satisfied:
- $f(\sigma_1, \ldots, \sigma_n) = 1$,
- the special decomposition $d_n S(f)$ is generated by the function $f(x_1, \ldots, x_n)$,
- the special decomposition $d_n S(f)G$ is generated by the special decomposition $d_n S(f)$ as a result of admissible changes under the assignment tuple $(\sigma_1, \ldots, \sigma_n)$,
- $h(x_1, \ldots, x_n)$ is generated by the special decomposition $d_n S(f)G$.

*Theorem* 4.2. Let $f(x_1, \ldots, x_n)$ be a Boolean function of $n$ variables represented in conjunctive normal form, and let for some assignment tuple $(\sigma_1, \ldots, \sigma_n)$, $f(\sigma_1, \ldots, \sigma_n) = 1$.

If the function $h(x_1, \ldots, x_n)$ is generated by the function $f(x_1, \ldots, x_n)$ as a result of admissible changes under the assignment tuple $(\sigma_1, \ldots, \sigma_n)$, then $h(x_1, \ldots, x_n)$ is also a satisfiable function.

*Proof.* Suppose that $d_n S(f)$ is a special decomposition of the set $S(f)$ generated by the function $f(x_1, \ldots, x_n)$. Since $f(x_1, \ldots, x_n)$ is a satisfiable function and $f(\sigma_1, \ldots, \sigma_n) = 1$, then according to Theorem 2.7 the ordered set
$$c_n S(f) = \{F_1^{\sigma_1}, F_2^{\sigma_2}, \ldots, F_n^{\sigma_n}\}$$
is a special covering for the set $S(f)$ under the special decomposition $d_n S(f)$.

The function $h(x_1, \ldots, x_n)$ is generated by the function $f(x_1, \ldots, x_n)$ by the admissible changes under the assignment tuple $\sigma_1, \ldots, \sigma_n)$. This means that:
- the special decomposition $d_n S(f)$ generates the special decomposition $d_n S(f)G$ as a result of admissible changes under the assignment tuple $(\sigma_1, \ldots, \sigma_n)$,
- the decomposition $d_n S(f)G$ generates the function $h(x_1, \ldots, x_n)$.

According to Theorem 3.3, there is a special covering for the set $S(f)$ under the special decomposition $d_n S(f)G$. But then, according to Theorem 2.5, the function $h(x_1, \ldots, x_n)$ is satisfiable. ∇

*Corollary* 4.2.1. Let $f(x_1, \ldots, x_n)$ be a satisfiable function of $n$ variables represented in conjunctive normal form.

If the tuple $(\sigma_1, \ldots, \sigma_n)$ is a satisfying assignment for the function $f(x_1, \ldots, x_n)$, and the function $h(x_1, \ldots, x_n)$ is generated by the function $f(x_1, \ldots, x_n)$ as a result of admissible changes under the assignment tuple $(\sigma_1, \ldots, \sigma_n)$, then $h(\sigma_1, \ldots, \sigma_n) = 1$.

*Proof.* It is easy to see that during the admissible changes in the decomposition $d_n S(f)$ only the contents of some subsets included in $c_n S(f)$ are changed, but in general the special covering does not lose elements. This means that the subsets with superscripts $\sigma_1, \ldots, \sigma_n$, respectively, cover the set $S(f)$. Therefore, $h(\sigma_1, \ldots, \sigma_n) = 1$. ▽

Let's explore the nature of the concept of function generation by a function.

For the function $f(x_1, \ldots, x_n)$ and for the satisfying assignment $(\sigma_1, \ldots, \sigma_n)$ we define the class of functions, denoted as $Gf[\sigma_1, \ldots, \sigma_n]$, as follows:

(1) $f(x_1, \ldots, x_n) \in Gf[\sigma_1, \ldots, \sigma_n]$.

(2) if the function $h_2(x_1, \ldots, x_n)$ is generated by the function $h_1(x_1, \ldots, x_n)$ as a result of admissible changes under the assignment tuple $(\sigma_1, \ldots, \sigma_n)$, then, if $h_1(x_1, \ldots, x_n) \in Gf[\sigma_1, \ldots, \sigma_n]$ then $h_2(x_1, \ldots, x_n) \in Gf[\sigma_1, \ldots, \sigma_n]$.

(3) the class $Gf$ contains only functions satisfying conditions (1) and (2).

*Theorem* 4.3. Let $f(x_1, \ldots, x_n)$ and $h(x_1, \ldots, x_n)$ be Boolean functions of $n$ variables and each of them is represented in conjunctive normal form with $m$ clauses.

If there exists a Boolean satisfiable tuple $(\sigma_1, \ldots, \sigma_n)$ such that
$$f(\sigma_1, \ldots, \sigma_n) = 1 \text{ and } h(\sigma_1, \ldots, \sigma_n) = 1,$$
then $h(x_1, \ldots, x_n) \in Gf(\sigma_1, \ldots, \sigma_n)$ and $f(x_1, \ldots, x_n) \in Gh(\sigma_1, \ldots, \sigma_n)$.

*Proof.* Let $S(f) = \{c_1, \ldots, c_n\}$ be the ordered set of clauses of the function $f(x_1, \ldots, x_n)$, and let $S(h) = \{e_1, \ldots, e_n\}$ be the ordered set of clauses of the function $h(x_1, \ldots, x_n)$.

According to Lemma 2.2 the functions $f(x_1, \ldots, x_n)$ and $h(x_1, \ldots, x_n)$ generate special decompositions $d_n S(f)$ and $d_n S(h)$ of the sets $S(f)$ and $S(h)$, respectively:
$$d_n S(f) = \{(F_1^0, F_1^1), (F_2^0, F_2^1), \ldots, (F_n^0, F_n^1)\},$$
$$d_n S(h) = \{(H_1^0, H_1^1), (H_2^0, H_2^1), \ldots, (H_n^0, H_n^1)\}.$$

In addition, the ordered set
$$c_n S(f) = \{F_1^{\sigma_1}, F_2^{\sigma_2}, \ldots, F_n^{\sigma_n}\}$$
is a special covering for the set $S(f)$ under the decomposition $d_n S(f)$. Also, the ordered set
$$c_n S(h) = \{H_1^{\sigma_1}, H_2^{\sigma_2}, \ldots, H_n^{\sigma_n}\}$$
is a special covering for the set $S(h)$ under the decomposition $d_n S(h)$.

Thus, we will proof that the special decomposition $d_n S(h)$ is generated by the special decomposition $d_n S(f)$ as a result of admissible changes under the superscript tuple $(\sigma_1, \ldots, \sigma_n)$.

Also, the special decomposition $d_n S(f)$ is generated by the special decomposition $d_n S(h)$ as a result of admissible changes under superscript tuple $(\sigma_1, \ldots, \sigma_n)$.

Let for some $i \in \{1, \ldots, n\}$,
$$F_i^{\sigma_i} = \{c_{i_1}, \ldots, c_{i_p}\} \text{ and } H_i^{\sigma_i} = \{e_{j_1}, \ldots, e_{j_q}\}.$$

By the definition of these subsets,
$$x_i^{\sigma_i} \in c_{i_k} \text{ for any } c_{i_k} \in \{c_{i_1}, \ldots, c_{i_p}\},$$
$$x_i^{\sigma_i} \in e_{j_k} \text{ for any } e_{j_k} \in \{e_{j_1}, \ldots, e_{j_q}\}.$$

Let's describe a procedure for obtaining the subset $H_i^{\sigma_i}$ instead the subset $F_i^{\sigma_i}$ as a result of admissible changes in the decomposition $d_n S(f)$.

The procedure consists of applying the points of Definition 3.1 to the subsets included in the decomposition $d_n S(f)$.

We will assume that none of the subsets included in the set $c_n S(f)$ is empty, otherwise we can add an element to this subset according to admissible changes. This does not affect the estimation of the complexity of the entire procedure.

a) We sequentially remove all elements from subsets not included in the set $c_n S(f)$. As a result, any ordered pair $(F_i^0, F_i^1)$ will take the form

$$(F_i^0, \emptyset), \text{ if } F_i^0 \in c_n S(f) \text{ or } (\emptyset, F_i^1), \text{ if } F_i^1 \in c_n S(f).$$

We can do this applying the point (i.1) of the Definition 3.1.

It is easy to see, that as a result of these operations some clauses of the function are change, so the function is changed.

At the same time, the resulting function is satisfiable since all changes are made in accordance with the admissible changes under the same assignment tuple.

b) let's consider the following cases for the clauses of the subsets

$$F_i^{\sigma_i} = \{c_{i_1}, \ldots, c_{i_p}\} \text{ and } H_i^{\sigma_i} = \{e_{j_1}, \ldots, e_{j_q}\}.$$

Recall that our goal is to obtain the clauses of the set $H_i^{\sigma_i}$ instead of the set $F_i^{\sigma_i}$ in the decomposition $d_n S(f)$ using the admissible changes.

b.1) suppose that $p = q$.

In this case, for any number $i_k \in \{i_1, \ldots, i_p\}$, we proceed as follows:

We compare the pairs of clauses $c_{i_k}$ and $e_{i_k}$:

- if these clauses are the same, we consider $e_{i_k}$ as a clause of the subset $F_i^{\sigma_i}$ and compare other pairs.

- let these clauses be different. That is, there is a literal, let it be $x_s^{\alpha_s}$, such that
$$x_s^{\alpha_s} \in e_{i_k} \text{ and } x_s^{\alpha_s} \notin c_{i_k}.$$

In this case, we add $x_s^{\alpha_s}$ to the clause $c_{i_k}$.

Recall that adding the literal $x_s^{\alpha_s}$ to the clause $c_{i_k}$ means adding the clause $c_{i_k}$ to the subset $F_s^{\alpha_s}$. According to the point (i.2), of the Definition 3.1, this is possible if $c_{i_k} \notin F_s^{\overline{\alpha}_s}$.

To show that the clause $c_{i_k}$ can be added to the subset $F_s^{\alpha_s}$, consider two cases:

- $F_s^{\alpha_s} \in c_n S(f)$. In this case $F_s^{\overline{\alpha}_s} = \emptyset$, therefore we can add $c_{i_k}$ to the subset $F_s^{\alpha_s}$ which will mean that the literal $x_s^{\alpha_s}$ is added to the clause $c_{i_k}$.

- $F_s^{\alpha_s} \notin c_n S(f)$. Then $F_s^{\overline{\alpha}_s} \in c_n S(f)$ and $F_s^{\alpha_s} = \emptyset$. So, we will add $c_{i_k}$ to the empty subset $F_s^{\alpha_s}$. In this case, if $c_{i_k}$ is included in $F_s^{\overline{\alpha}_s}$, we can remove it in accordance to the point (i.3) of the Definition 3.1, since $c_{i_k}$ is also included in another subset $F_i^{\sigma_i}$ of the set $c_n S(f)$.

Thus, in case b.1) by means of admissible changes, we can add all clauses included in the subset $H_i^{\sigma_i}$ to the subset $F_i^{\sigma_i}$.

b.2) if $p < q$, then we use the point (i.3) and add clauses to the subset $F_i^{\sigma_i}$ such that the number of clauses in it will be equal to the number of clauses in the subset $H_i^{\sigma_i}$.

As a result, we will get the case b.1).

b.3) if $p > q$, then using the point (i.3), we move some clauses from the subset $F_i^{\sigma_i}$ to other subsets such that the number of clauses in it will be equal to the number of clauses in the subset $H_i^{\sigma_i}$. As a result, we again get the case b.1).

c) after adding all the literals of the clause $e_{i_k}$ to the clause $c_{i_k}$, we proceed to remove from the clause $c_{i_k}$ literals that are not included in the clause $e_{i_k}$, as follows:

Let $x_r^{\alpha_r} \in c_{i_k}$ and $x_r^{\alpha_r} \notin e_{i_k}$.

Removing $x_r^{\alpha_r}$ from $c_{i_k}$ means removing $c_{i_k}$ from the subset $F_r^{\alpha_r}$. Note that we can do this using the point (i.3) of the Definition 3.1, since $c_{i_k}$ is also included in the subset $F_i^{\sigma_i}$.

Repeating the procedure according to described points for all $i \in \{1, \ldots, n\}$, we obtain the set $c_n S(h)$ instead of the set $c_n S(f)$.

d) by applying the point (i.2) we do the following:

For any subset $H_i^{\overline{\sigma}_i}$, which is not included in the special covering $c_n S(h)$, all clauses included in it are sequentially added to the subset $F_i^{\overline{\sigma}_i}$.

It is easy to see, that as a result, we obtain the special decomposition $d_n S(h)$. Therefore, we can assert that first side of the theorem is valid:
$$h(x_1, \ldots, x_n) \in Gf(\sigma_1, \ldots, \sigma_n).$$
Similarly, we can proof that $f(x_1, \ldots, x_n) \in Gh(\sigma_1, \ldots, \sigma_n)$. $\nabla$

Obviously, as a result of any step of the described procedure we obtain a new special decomposition of the set $S(f)$ and a new special covering for the set $S(f)$ under the obtained decomposition. Then, as a result of any step of the procedure, a satisfiable function is generated.

Applying the admissible changes to the subsets included in the special decomposition $d_n S(f)$, actually means performing the following operations with the clauses of the function $f$:

- the clause $c_j$ is removed from the subset $F_i^\alpha$. This means the removing of the literal $x_i^\alpha$ from the clause $c_j$,

- the clause $c_j$ is added to the subset $F_i^\alpha$. This means adding the literal $x_i^\alpha$ to the clause $c_j$,

- the clause $c_j$ is moved from the subset $F_i^\alpha$, to the subset, $F_j^\delta$. This means remove the literal $x_i^\alpha$ from the clause $c_j$ and add the literal $x_j^\delta$ to the obtained clause.

Thus, adding a literal to a certain clause, removing a literal from the certain clause or changing a literal with another literal according to conditions of admissible changes, we obtain a satisfiable function.

Using the theorems 4.2 and 4.3, it is easy to proof the equivalence theorem.

First, let's define a binary relation over the Boolean functions of $n$ variables and represented in conjunctive normal form with $m$ clauses. We will denote this relation by $G[\sigma_1, \ldots, \sigma_n]$ where $(\sigma_1, \ldots, \sigma_n)$ is a Boolean assignment tuple.

Suppose that $f(x_1, \ldots, x_n)$ and $h(x_1, \ldots, x_n)$ are Boolean function of $n$ variables, both in conjunctive normal form with $m$ clauses.

The ordered pair of these functions will be denoted as $(f, h)$.

We say that $(f, h) \in G[\sigma_1, \ldots, \sigma_n]$, if the following conditions are satisfied:

- $f(\sigma_1, \ldots, \sigma_n) = 1$,

- the function $h(x_1, \ldots, x_n)$ is generated by the function $f(x_1, \ldots, x_n)$ as a result of admissible changes under the satisfying assignment $(\sigma_1, \ldots, \sigma_n)$.

*Theorem* 4.4. For any Boolean assignment $(\sigma_1, \ldots, \sigma_n)$, the relation $G[\sigma_1, \ldots, \sigma_n]$ is an equivalence relation over the satisfiable Boolean functions of $n$ variables represented in conjunctive normal form with $m$ clauses.

*Proof.* It is easy to see that any satisfiable Boolean function represented in conjunctive normal form generates itself. So, if $f(\sigma_1, \ldots, \sigma_n) = 1$, then $(f, f) \in G[\sigma_1, \ldots, \sigma_n]$.

That is $GF[\sigma_1, \ldots, \sigma_n]$ is a reflexive relation.

Let's show that $G[\sigma_1, \ldots, \sigma_n]$ is a symmetric relation.

Suppose that the functions $f(x_1, \ldots, x_n)$ and $h(x_1, \ldots, x_n)$ are Boolean function of $n$ variables and both represented in conjunctive normal form with $m$ clauses such that
$$(f, h) \in G[\sigma_1, \ldots, \sigma_n].$$

This means that $f(\sigma_1, \ldots, \sigma_n) = 1$ and the function $h(x_1, \ldots, x_n)$ is generated by the function $f(x_1, \ldots, x_n)$ by admissible changes under the satisfying assignment $(\sigma_1, \ldots, \sigma_n)$.

According to Theorem 4.2 and Corollary 4.2.1, the function $h(x_1, \ldots, x_n)$ is satisfiable and $h(\sigma_1, \ldots, \sigma_n) = 1$. Obviously, the conditions of Theorem 4.3 are satisfied, therefore
$$f(x_1, \ldots, x_n) \in Gh(\sigma_1, \ldots, \sigma_n).$$

That is, the function $f(x_1, \ldots, x_n)$ is generated by the function $h(x_1, \ldots, x_n)$ as a result of admissible changes under the satisfying assignment $(\sigma_1, \ldots, \sigma_n)$, and so, $G[\sigma_1, \ldots, \sigma_n]$ is a symmetric relation.

Now let's prove that $G[\sigma_1, \ldots, \sigma_n]$ is a transitive relation.

Suppose that $f(x_1, \ldots, x_n)$, $g(x_1, \ldots, x_n)$ and $h(x_1, \ldots, x_n)$ are Boolean functions of $n$ variables and all represented in conjunctive normal form with $m$ clauses such that
$$(f, g) \in G[\sigma_1, \ldots, \sigma_n] \text{ and } (g, h) \in G[\sigma_1, \ldots, \sigma_n].$$

We will show that $(f, h) \in G[\sigma_1, \ldots, \sigma_n]$.

Actually, we will prove that the function $h(x_1, \ldots, x_n)$ is generated by the function $f(x_1, \ldots, x_n)$ as a result of admissible changes under the satisfying assignment $(\sigma_1, \ldots, \sigma_n)$.

Assume that as a result of admissible changes in the procedure for generating the function $g$, the clauses $c_{i_1}, \ldots, c_{i_p}$, belonging to the function $f$, become the clauses $c'_{i_1}, \ldots, c'_{i_p}$ of the function $g$.

Let also, as a result of admissible changes in the procedure for generating the function $h$, the clauses $c_{j_1}, \ldots, c_{j_q}$, belonging to the function $g$, become clauses $c'_{j_1}, \ldots, c'_{j_q}$ of the function $h$.

It is easy to notice, that any clause included in the function $h$ is either included in the function $f$ or is obtained from some clause of the function $f$ as a result of admissible changes.

Combining all changes carried out both in the procedure for generating function g and in the procedure for generating function h, we will get the procedure that generates the function $h$ from the function $f$.

We obtained that $(f, h) \in G[\sigma_1, \ldots, \sigma_n]$, that is $G[\sigma_1, \ldots, \sigma_n]$ is a transitive relation.

Thus, we proved that $G[\sigma_1, \ldots, \sigma_n]$ is an equivalence relation over the satisfiable functions of $n$ variables represented in conjunctive normal form with $m$ clauses. $\nabla$

## 5. *Complexity Estimations*

During the proofs of Theorems, we describe procedures that implement the proofs.

On the other hand, for any assignment tuple $(\sigma_1, \ldots, \sigma_n)$, a new function is generated as a result of any admissible steps under this tuple.

Therefore, an important issue is to estimate the complexity of the procedure for generating an arbitrary function from another arbitrary function.

So, let's move on to estimate the complexities of the described procedures.

In the previous sections we defined the following basic procedures:

(a) A procedure for generating a special decomposition of the set of clauses of the given Boolean function in conjunctive normal form.

(b) A procedure for generating a Boolean function in conjunctive normal form based on a given special decomposition of a certain set.

(c) A procedure for generating a satisfiable function from another satisfiable function by admissible changes under some satisfying assignment.

It is important to note that the procedures (a) and (b) are involved in the procedure (c).

*Data Representations*

We are dealing with Boolean functions represented in conjunctive normal form.

Further, for our purposes, it is technically convenient to use the matrices with the elements 0, -1, 1, to represent these functions as follows:

The rows of the matrix will represent the clauses of the function.

The none-zero elements of the rows will represent the literals included in the clauses.

Let we are given a function $f(x_1, \ldots, x_n)$ of $n$ variables in conjunctive normal form with $m$ clauses. We will assume that the clauses of this function are numbered in an arbitrary order, and $1, \ldots, m$ are their numbers. Let them be $c_1, \ldots, c_m$.

For a function $f$ we form an $(m \times n)$ matrix, denoted by $(f)\text{cnf}$, as follows:

$(f)\text{cnf}(i, j) = -1$, if the negative literal $\bar{x}_j$ is included in the clause $c_i$,

$(f)\text{cnf}(i, j) = 0$, if none of the literals $x_j$ and $\bar{x}_j$ is included in the clause $c_i$,

$(f)\text{cnf}(i, j) = 1$, if the positive literal $x_j$ is included in the clause $c_i$.

Obviously, for any $i \in \{1, \ldots, n\}$, the $i$-th row of the matrix $(f)\text{cnf}$ is uniquely determined by the clause $c_i$ of the function.

Also, for any $i \in \{1, \ldots, n\}$, the clause $c_i$ of the function is uniquely determined by the $i$-th row of the corresponding matrix $(f)\text{cnf}$.

So, any Boolean function $f(x_1, \ldots, x_n)$ represented in conjunctive normal form is uniquely determined by the corresponding matrix $(f)\text{cnf}$, and the matrix $(f)\text{cnf}$ is uniquely determined by the Boolean function $f(x_1, \ldots, x_n)$ in conjunctive normal form.

W*e will say that the Boolean function $f(x_1, \ldots, x_n)$ of $n$ variables in conjunctive normal form with $m$ clauses is represented by the $(m \times n)$-matrix $(f)$*cnf.

It is easy to see, that an $(m \times n)$-matrix with elements 0, -1, 1 only, represents a Boolean function if and only if it does not contain a row with only zeros and a column with only zeros.

Note that the row with only zeros means that the function contains an empty clause, which cannot be satisfiable. Therefore, it makes no sense to consider such a function.

In addition, a column with only zeros means that the corresponding variable is not included in any clause. In this case, we will not consider this function to be a function of $n$ variables.

We will also be dealing with a non-empty set of $m$ elements, denoted as
$$S = \{e_1, \ldots, e_m\}.$$
Let $d_n S$ be a special decomposition of the set $S$:
$$d_n S = \{(M_1^0, M_1^1), \ldots, (M_n^0, M_n^1)\}.$$

We will consider $S$ as an ordered set, assuming that its elements are numbered in the order in which they appear in the set notation. It will not lead to any ambiguity.

Similar to the case of Boolean functions, it is technically convenient to represent the special decomposition of the set $S$ using an ordered pair of (0,1)-matrices.

Based on a special decomposition we form two $(n \times m)$-matrices, denoted $sM0$ and $sM1$, respectively. The elements of these matrices and are determined as follows:

For $i \in \{1, \ldots, n\}$ and $j \in \{1, \ldots, m\}$,

$$sM0(i,j) = \begin{cases} 0, & if\ e_j \notin M_i^0 \\ 1, & if\ e_j \in M_i^0 \end{cases} \qquad sM1(i,j) = \begin{cases} 0, & if\ e_j \notin M_i^1 \\ 1, & if\ e_j \in M_i^1 \end{cases}.$$

*We will say that the ordered pair of matrices $(sM0, sM1)$ corresponds to the special decomposition $d_nS$ if this pair is formed in described manner on the basis of the decomposition $d_nS$.*

On the other hand, the special decomposition $d_nS$ is determined by the corresponding ordered pair of matrices $(sM0, sM1)$ as follows:

It is easy to see, that the elements of the subsets $M_i^0$ and $M_i^1$ are uniquely determined by the 1s of the $i$-th rows of the matrices $sM0$ and $sM1$, respectively:

For any $i \in \{1, \ldots, n\}$, $M_i^\alpha = \{e_j \in S\ /\ sM0(i,j) = 1\}$ and
$$M_i^{1-\alpha} = \{e_j \in S\ /\ sM1\ (i,j) = 1\},$$

This means that any row of the matrices $sM0$ and $sM1$ corresponds to some subset included in the ordered pairs of a special decomposition. And also, any subset included in an ordered pair is uniquely determined by corresponding row of one of these matrices.

The ordered pair $(M_i^0, M_i^1)$ of the decomposition $d_nS$ will be determined by ordered pair of $i$-th rows of the matrices $sM0$ and $sM1$.

Obviously, a pair of (0,1)-matrices correspond to any special decomposition of a set.

In addition, all parameters of the special decomposition are uniquely determined by the corresponding pair of (0, 1)-matrices.

*We will say that the pair of (0,1)-matrices $(sM0, sM1)$ is generated by the Boolean function $f(x_1, \ldots, x_n)$, if this pair corresponds to the special decomposition $d_nS(f)$.*

Recall that for any $i \in \{1, \ldots, n\}$ the subsets $F_i^0$ and $F_i^1$ are composed as follows:

$F_i^0 = \{c_j\ /\ c_j \in S(f)$ and $c_j$ contains the literal $\bar{x}_i$, $(j \in \{1, \ldots, m\})\}$,

$F_i^1 = \{c_j\ /\ c_j \in S(f)$ and $c_j$ contains the literal $x_i$, $(j \in \{1, \ldots, m\})\}$.

$c_j$ is a $j$-th clause of the function $f(x_1, \ldots, x_n)$.

According to Lemma 2.2, the ordered set of the ordered pairs of these subsets compose the special decomposition $d_nS(f)$.

Let's denote by $((f)sM0, (f)sM1)$ the pair of (0,1)-matrices, which is formed as follows:

$$(f)sM0(i,j) = \begin{cases} 0, & if\ c_j \notin F_i^0 \\ 1, & if\ c_j \in F_i^0 \end{cases} \qquad (f)sM1(i,j) = \begin{cases} 0, & if\ c_j \notin F_i^1 \\ 1, & if\ c_j \in F_i^1 \end{cases}.$$

We will say that the pair of (0,1)-matrices $((f)sM0, (f)sM1)$ corresponds to the special decomposition $d_nS(f)$.

On the other hand, $c_j \in F_i^\alpha$ if the literal $x_i^\alpha$ is included in the clause $c_j$. So,

$$(f)sM0(i,j) = \begin{cases} 0, & if\ x_i^0 \notin c_j \\ 1, & if\ x_i^0 \in c_j \end{cases} \qquad (f)sM1(i,j) = \begin{cases} 0, & if\ x_i \notin c_j \\ 1, & if\ x_i \in c_j \end{cases}.$$

In addition, we will use the following notation for any $i \in \{1, \ldots, n\}$:

$$(f)M0(i) = ((f)sM0(i, 1), \ldots, (f)sM0(i, m))$$
$$(f)M1(i) = ((f)sM1(i, 1), \ldots, (f)sM1(i, m)).$$

It is obvious, that the ordered pair $((f)M0(i), (f)M1(i))$ is the ordered pair of $i$-th rows of the matrices $(f)sM0$ and $(f)sM1$, respectively.

*The Complexity of the Described Procedures.*

Let we are given a special decomposition $d_nS$, and let $(sM0, sM1)$ be a pair of (0,1)-matrices corresponding to this special decomposition.

Also, let we are given a Boolean function $f(x_1, \ldots, x_n)$ in conjunctive normal form, and let $(f)$cnf be the corresponding matrix with the elements 0, -1 and 1.

We often identify a special decomposition of a nonempty set of $m$ elements containing $n$ ordered pairs of subsets with the corresponding pair of (0,1)-matrices of the size $(n \times m)$.

We will also often identify a Boolean function of $n$ variables, in conjunctive normal form with $m$ clauses, with the corresponding $(m \times n)$-matrix with the elements 0, -1 and 1.

*Definition* 5.1. (i) The total number of 1s in the pair of matrices $(sM0, sM1)$ will be called the number of the input data of the special decomposition $d_nS$.

(ii) The total number of non-zero elements included in the matrix $(f)$cnf will be called the length of input data of the function $f(x_1, \ldots, x_n)$.

It is easy to see that the input data of the special decomposition is actually the total number of elements included in all subsets that make up this decomposition.

In addition, the length of input data of the function $f(x_1, \ldots, x_n)$ is actually the total number of literals in all clauses of the function.

*Definition* 5.2. The following operations will be called elementary:

- assigning a value to a function variable or assigning a value to an array element,
- addition and subtraction of numbers,
- comparison of two numbers,
- recognition of a literal.

*Proposition* 5.3. Let $f(x_1, \ldots, x_n)$ be a Boolean function of $n$ variables represented in conjunctive normal form with $m$ clauses.

The number of elementary operations required to obtain the pair of (0,1)-matrices $((f)sM0, (f)sM1)$ does not exceed the number $c \times n \times m$ for some constant $c$.

*Proof.* We will form the matrices $(f)sM0$ and $(f)sM1$ based on the matrix $(f)$cnf corresponding to the conjunctive normal form of the function $f$, and using the formulas described in the previous section.

That is, we sequentially consider all the elements of any row of the matrix $(f)$cnf and form the corresponding rows of the matrices $(f)sM0$ and $(f)sM1$.

Recall that this corresponds to considering all the literals of any clause.
Let's immediately describe the algorithm on how to do this.

```
for j = 1 to n do:
    for i = 1 to m do:
            if (f)cnf(j, i) = 1:
            sMā(i, j) := 1
            sMα(i, j) := 0
        elif (f)cnf(j, i) =-1:
            sMα(i, j) := 1
            sMā(i, j) := 0
        else:
            (sMā(i, j) := 0)
            (sMα(i, j) := 0)
        endif;
    endfor i;
endfor j;
```

It is easy to see, that all operations in the described procedure are elementary, and as a result of the procedure, the matrices $(f)sM0$ and $(f)sM1$ are formed correctly.

Also, the number of elementary operations required to perform the procedure does not exceed the number $c \times (n \times m)$ for some constant $c$. ∇

*Proposition* 5.4. Let we are given an ordered set $S = \{e_1, \ldots, e_m\}$, and
$$d_n S = \{(M_1^0, M_1^1), \ldots, (M_i^0, M_i^1), \ldots, (M_n^0, M_n^1)\},$$
is a special decomposition of the set $S$.

The number of elementary operations required to obtain the Boolean function generated by the special decomposition $d_n S$ does not exceed the number $c \times n \times m$ for some constant $c$.

*Proof.* Let's denote by $h(x_1, \ldots, x_n)$ the function which is generated by the special decomposition $d_n S$. To obtain this function, we form the matrix $(h)$cnf, which will correspond to the function $h(x_1, \ldots, x_n)$.

We will use the procedure described in the section 2.4.

Suppose that $(sM0, sM\overline{1})$ is the ordered pair of (0,1)-matrices corresponding to the special decomposition $d_n S$. Recall that
$$sM0(i, j) = 1, \text{ if } e_j \in M_i^0 \text{ and } sM1(i, j) = 1, \text{ if } e_j \in M_i^1.$$

We proceed as follows:

Based on the positions of the element $e_j \in S$ in the decomposition $d_n S$, we form the row of $m$ elements consisting by zeros and ones, which will be the $j$-th row of the matrix $(h)$cnf.

Thus, for any $i \in \{1, \ldots, n\}$, we consider the $i$-th row of the matrix $sM0$ and $i$-th row of the matrix $sM1$, and do the following:

If $sM0(i, j) = 1$ for some $j \in \{1, \ldots, m\}$, then $(h)$cnf$(j, i) = -1$,
If $sM1(i, j) = 1$ for some $j \in \{1, \ldots, m\}$, then $(h)$cnf$(j, i) = 1$.
The matrix will be formed as a result of the following algorithm:

```
for j = 1 to m do:
    for i = 1 to n do:
        if sM0(i, j) = 1:
```

```
            (h)cnf(j, i) = -1
    elif sM1(i, j) = 1:
            (h)cnf(j, i) = 1
    else:
            (h)cnf(j, i) = 0
    endif;
  endfor i;
endfor j;
```

It is easy to see, that all operations in the described procedure are elementary, and as a result of the procedure, the matrix $(h)\mathrm{cnf}(j, i)$ is formed correctly.

Also, the number of elementary operations required to perform the procedure does not exceed the number $c \times n \times m$ for some constant $c$. ∇

The Propositions 5.3 and 5.4 prove the following:

*Any Boolean function generates a special decomposition in polynomial time, and any special decomposition of a set generates a Boolean function in conjunctive normal form in polynomial time.*

Comparing these results with Theorems 2.3 and 2.5, we find:

*The problem of finding a special covering for a set and Boolean satisfiability problem are polynomially equivalent.*

Therefore, it is easy to see that:

*The problem of finding a special covering for a set is an NP-complete problem.*

This result has been proven in more detail in [8].

*The Complexity of the Generating Procedure*

<u>Theorem</u> 5.5. Let $f(x_1, \ldots, x_n)$ be a Boolean function represented in conjunctive normal form with $m$ clauses, and let $(\sigma_1, \ldots, \sigma_n)$ be satisfiable assignment tuple for this function,
$$f(\sigma_1, \ldots, \sigma_n) = 1.$$

If $h(x_1, \ldots, x_n)$ is a Boolean function generated by the function $f(x_1, \ldots, x_n)$ as a result of admissible changes under the assignment tuple $(\sigma_1, \ldots, \sigma_n)$, then:

The number of elementary operations required to perform this generating procedure does not exceed the number $c \times n \times m$ for some constant $c$.

<u>Proof</u>. Since the function $h(x_1, \ldots, x_n)$ is generated by the function $f(x_1, \ldots, x_n)$ as a result of admissible changes under the tuple of satisfying assignment $(\sigma_1, \ldots, \sigma_n)$, then according to Corollary 4.2.1, $h(\sigma_1, \ldots, \sigma_n) = 1$.

In addition, the following conditions are satisfied by the Definition 4.1:

- the function $f(x_1, \ldots, x_n)$ generates the special decomposition $d_n S(f)$ by Lemma 2.2,
- the special decomposition $d_n S(f)$ generates the special decomposition $d_n S(f)G$ as a result of admissible changes under the assignment tuple $(\sigma_1, \ldots, \sigma_n)$,
- the special decomposition $d_n S(f)G$ generates the function $h(x_1, \ldots, x_n)$.

Thus, it is enough to estimate the number of required elementary operations for each of these procedures.

According to Proposition 5.3, the number of elementary operations required to obtain the pair of (0,1)-matrices $((f)sM0, (f)sM1)$ generated by the function $f(x_1, \ldots, x_n)$ does not exceed the number $c \times n \times m$ for some constant $c$.

According to Proposition 5.4, the number of elementary operations required to obtain the Boolean function generated by an ordered pair of (0,1)-matrices $((f)sM0, (f)sM1)$, does not exceed the number $c \times n \times m$ for some constant $c$.

So, we need to estimate the number of elementary operations for generating the special decomposition $d_n S(f) G$ by the special decomposition $d_n S(f)$.

For convenience, here we will use $d_n S(h)$ instead of the notation $d_n S(f) G$ for the decomposition that generate the function $h(x_1, \ldots, x_n)$.

We will use the pair of (0,1)-matrices $((f)sM0, (f)sM1)$ and $((h)sM0, (h)sM1)$ that correspond to the special decompositions $d_n S(f)$ and $d_n S(h)$, respectively.

Since the subsets $F_i^{\sigma_i}$ and $H_i^{\sigma_i}$ correspond to the rows $(f)M\sigma_i(i)$ and $(h)M\sigma_i(i)$, respectively, then the ordered sets $c_n S(f)$ and $c_n S(h)$,
$$c_n S(f) = \{F_1^{\sigma_1}, \ldots, F_i^{\sigma_i}, \ldots, F_n^{\sigma_n}\},$$
$$c_n S(h) = \{H_1^{\sigma_1}, \ldots, H_i^{\sigma_i}, \ldots, H_n^{\sigma_n}\},$$
correspond to the following ordered sets of the rows:
$$\{(f)M\sigma_1(1), (f)M\sigma_2(2), \ldots, (f)M\sigma_n(n)\},$$
$$\{(h)M\sigma_1(1), (h)M\sigma_1(1), \ldots, (h)M\sigma_n(n)\},$$
respectively.

Thus, we will estimate the maximum number of elementary operations required to generate the pair of (0,1)-matrices $((h)sM0, (h)sM1)$ based on the pair of (0,1)-matrices $((f)sM0, (f)sM1)$ using admissible changes under the tuple $(\sigma_1, \ldots, \sigma_n)$.

We will use the procedure similar to the procedure described in the Theorem 4.3.

Let's note that we will consider the case when none of the rows included in the set
$$\{(f)M\sigma_1(1), (f)M\sigma_2(2), \ldots, (f)M\sigma_n(n)\}$$
does not consist only of zeros, otherwise, we add 1 to this row in accordance with admissible changes. This does not affect the estimation of the complexity of the entire procedure.

The procedure described in the Theorem 4.3 actually consists of the following points:

a) removal of all subsets that are not included in the special covering $c_n S(f)$.

This means sequentially assign zeros to all elements of the rows
$$(f)M\overline{\sigma}_1(1), (f)M\overline{\sigma}_2(2), \ldots, (f)M\overline{\sigma}_n(n).$$

That is, the elements of any row not included in the set corresponding to the special covering $c_n S(f)$ are assigned zero. Since the number of elements of any row does not exceed $m$, then the number of elementary operations for this point does not exceed the number $(n \times m)$.

b) for any $i \in \{1, \ldots, n\}$ we add all clauses included in the subset $H_i^{\sigma_i}$ to the subset $F_i^{\sigma_i}$.

Therefore, for any $i \in \{1, \ldots, n\}$, we compare the elements of the rows $(f)M\sigma_i(i)$ and $(h)M\sigma_i(i)$ corresponding to the subsets $F_i^{\sigma_i}$ and $H_i^{\sigma_i}$, respectively,
$$(f)M\sigma_i(i) = \{(f)sM\sigma_i(i, 1), \ldots, (f)sM\sigma_i(i, m)\},$$
$$(h)M\sigma_i(i) = \{(h)sM\sigma_i(i, 1), \ldots, (h)sM\sigma_i(i, m)\},$$
and proceed as follows:

For any $j \in \{1, \ldots, m\}$, we assign $(f)sM\sigma_i(i, j) = 1$, if $(h)sM\sigma_i(i, j) = 1$.

Obviously, the number of elementary operations for this point also does not exceed the number $(n \times m)$.

c) for any $i \in \{1, \ldots, n\}$, all clauses included in the subset $F_i^{\sigma_i}$ and not included in $H_i^{\sigma_i}$ will be removed from the subset $F_i^{\sigma_i}$. That is, if for some $j \in \{1, \ldots, m\}$,
$$(f)sM\sigma_i(i, j) = 1 \text{ and } (h)sM\sigma_i(i, j) = 0,$$
then we will assign $(f)sM\sigma_i(i, j) = 0$, which means removing the j-th clause from the subset $F_i^{\sigma_i}$.

Let's show that as a result of this removal, the special covering $c_n S(h)$ will not lose an element. Since the ordered set of rows
$$\{(h)M\sigma_1(1), (h)M\sigma_2(2), \ldots, (h)M\sigma_n(n)\}$$
corresponds to the special covering $c_n S(h)$, then definitely $(h)sM\sigma_i(i, j) = 1$ for some
$$i \in \{1, \ldots, n\} \text{ and } j \in \{1, \ldots, m\}.$$

Also, by point b) we have $(f)sM\sigma_i(i, j) = 1$ for any $i$ and $j$ such that $(h)sM\sigma_i(i, j) = 1$.

This means that the operation of assigning $(f)sM\sigma_i(i, j) = 0$ corresponds to an admissible change, since for the same value of $j$ and for another value of $i$, $(f)sM\sigma_i(i, j) = 1$.

Thus, by this point the procedure performs the following operations:

For any $i \in \{1, \ldots, n\}$, it runs over the $i$-th row of the matrix $(h)sM\sigma_i$ and considers the values of its elements.
$$(h)M\sigma_i(i) = \{(h)sM\sigma_i(i, 1), \ldots, (h)sM\sigma_i(i, m)\}.$$

If it turns out that
$$(h)sM\sigma_i(i, j) = 0 \text{ and } (f)sM\sigma_i(i, j) = 1$$
for some $i \in \{1, \ldots, n\}$ and $j \in \{1, \ldots, m\}$, then the element $(f)sM\sigma_i(i, j)$ of the matrix $(f)sM\sigma_i$ is assigned the value 0.

It is easy to see that the described procedure requires no more than $c \times n \times m$ elementary operations for some constant $c$.

d) for any pair of rows $(f)M\overline{\sigma}_i(i)$ and $(h)M\overline{\sigma}_i(i)$,
$$(f)M\overline{\sigma}_i(i) \in \{(f)M\overline{\sigma}_1(1), (f)M\overline{\sigma}_2(2), \ldots, (f)M\overline{\sigma}_n(n)\},$$
$$(h)M\overline{\sigma}_i(i) \in \{(h)M\overline{\sigma}_1(1), (h)M\overline{\sigma}_2(2), \ldots, (h)M\overline{\sigma}_n(n)\},$$
the elements of the row $(f)M\overline{\sigma}_i(i)$ are assigned by the corresponding elements of the row $(h)M\overline{\sigma}_i(i)$. It is easy to see, that the procedure for performing this point requires no more than $(n \times m)$ elementary operation.

Thus, as a result of the procedures described in points a), b), c) and d) we obtain the pair of (0,1)-matrices $((h)sM0, (h)sM1)$ based on the pair of (0,1)-matrices $((f)sM0, (f)sM1)$ using admissible changes under the assignment tuple $(\sigma_1, \ldots, \sigma_n)$.

Obviously, the number of elementary operations for all described procedures does not exceed the number $c \times n \times m$ for some constant $c$. $\nabla$

Combining the results of the theorems 4.3 and 5.5, we can formulate the following:

*Theorem* 5.6. Let $f(x_1, \ldots, x_n)$ and $h(x_1, \ldots, x_n)$ be arbitrary Boolean functions of $n$ variables represented in conjunctive normal form with $m$ clauses.

If there is an assignment tuple $(\sigma_1, \ldots, \sigma_n)$ such that
$$f(\sigma_1, \ldots, \sigma_n) = 1 \text{ and } h(\sigma_1, \ldots, \sigma_n) = 1,$$
then, the function $f(x_1, \ldots, x_n)$ generates the function $h(x_1, \ldots, x_n)$ as a result of admissible changes under the assignment tuple $(\sigma_1, \ldots, \sigma_n)$ in no more than $c \times (n \times m)$ elementary operations, for some constant $c$.

*Proof.* The proof follows directly from the Theorems 4.3 and 5.5. $\nabla$

The following simple algorithm ensures that the function $h(x_1, \ldots, x_n)$ is generated by the function $f(x_1, \ldots, x_n)$ as a result of admissible changes. The algorithm uses ordered pairs $((f)sM0, (f)sM1)$ and $((h)sM0, (h)sM1)$, corresponding to special decompositions of the sets $S(f)$ and $S(h)$, as well as the satisfiable assignment tuple $(\sigma_1, \ldots, \sigma_n)$.

```
for i = 1 to n do:              // point (a) //
    for j = 1 to m do:
        (f)sM σ̄_i(i, j) = 0
    endfor j;
endfor i;

for i = 1 to n do:              // point (b) //
    for j = 1 to m do:
        if (h)sMσ_i(i, j) = 1:
            (f)sMσ_i(i, j) = 1
        endif;
    endfor j;
endfor i;

for i = 1 to n do:              //point (c) //
    for j = 1 to m do:
        if (f)sMσ_i(i, j) = 1 and (h)sMσ_i(i, j) = 0:
            (f)sMσ(i, j) = 0
        endif;
    endfor j;
endfor i;

for i = 1 to n do:              //point (d) //
    for j = 1 to m do:
        (f)sM σ̄_i(i, j) = (h)sM σ̄_i(i, j)
    endfor j;
endfor i;
```

It is easy to see that any operation performed by this algorithm is an admissible change, and therefore any operation generates a satisfiable function.

## 6. *Extension of Admissible Changes*

In this section, we will extend the concept of admissible changes by adding a new operation to the operations of Definition 3.1. In addition, we will explore extended admissible changes in special decompositions generated by Boolean functions.

Let we are given an ordered set $S = \{e_1, \ldots, e_m\}$, and
$$d_n S = \{(M_1^0, M_1^1), \ldots, (M_i^0, M_i^1), \ldots, (M_n^0, M_n^1)\},$$
is a special decomposition of the set $S$.

In previous sections we studied admissible changes that are done only by means of elements of different subsets.

Now, to the operations of admissible changes, an operation will be added that will deal with ordered pairs of special decomposition.

For any $i \in \{1, \ldots, n\}$, we will say that the ordered pair $(M_i^1, M_i^0)$ is obtained as a result of permutation the components of the ordered pair $(M_i^0, M_i^1)$.

*Definition* 6.1. Changes in the decomposition $d_n S$ are called extended admissible changes, if permutation of the components of some ordered pair included in the special decomposition is added to the operations of admissible changes.

It is easy to notice that adding a new operation to the operations of admissible changes actually means admitting *I*-thansformations in the special decomposition.

According to Lemma 1.4, this means that as a result of applying the new operation, the conditions of the special decomposition and special covering are preserved.

Let's consider the Boolean functions and the special decompositions generated by them.

Suppose that $f(x_1, \ldots, x_n)$ is a Boolean function represented in conjunctive normal form, and let the set
$$d_n S(f) = \{(F_1^0, F_1^1), \ldots, (F_i^0, F_i^1), \ldots, (F_n^0, F_n^1)\}$$
is a special decomposition of the set of clauses $S(f)$ of this function.

Recall that $(i_1, \ldots, i_k)I(d_n S(f))$ is a decomposition obtained as a result of permuting the components of the ordered pairs
$$\{(F_{i_1}^0, F_{i_1}^1), \ldots, (F_{i_k}^0, F_{i_k}^1)\}$$
in the special decomposition $d_n S(f)$.

According to Lemma 1.4, $(i_1, \ldots, i_k)I(d_n S(f))$ is a special decomposition of the set $S(f)$.

*Theorem* 6.2. Let for some $\{i_1, \ldots, i_k\} \subseteq \{1, \ldots, n\}$, the function $h(x_1, \ldots, x_n)$ be generated by the special decomposition $(i_1, \ldots, i_k)I(d_n S(f))$. Then:

$f(x_1, \ldots, x_n)$ is satisfiable function if and only if the function $h(x_1, \ldots, x_n)$ is satisfiable.

*Proof.* Let $f(x_1, \ldots, x_n)$ be a satisfiable function, and let the set of ordered pairs
$$d_n S(f) = \{(F_1^0, F_1^1), \ldots, (F_i^0, F_i^1), \ldots, (F_n^0, F_n^1)\}$$
be a special decomposition of the set $S(f)$.

Since $f(x_1, \ldots, x_n)$ is a satisfiable function, then According to Theorem 2.3, there is a special covering for the set $S(f)$ under the special decomposition $(d_n S(f))$. Let it be the set
$$c_n S(f) = \{F_1^{\alpha_1}, \ldots, F_i^{\alpha_i}, \ldots, F_n^{\alpha_n}\}.$$

The special decomposition $(i_1, \ldots, i_k)I(d_n S(f))$ is obtained as a result of permuting components of the ordered pairs
$$\{(F_{i_1}^0, F_{i_1}^1), \ldots, (F_{i_k}^0, F_{i_k}^1)\}$$
in the special decomposition $d_n S(f)$. So, by definition
$$(i_1, \ldots, i_k)I(d_n S(f)) = \{(F_1^{\sigma_1}, F_1^{\overline{\sigma}_1}), \ldots, (F_i^{\sigma_i}, F_i^{\overline{\sigma}_i}), \ldots, (F_n^{\sigma_n}, F_n^{\overline{\sigma}_n})\}$$

$$\text{for } \sigma_i = \begin{cases} 0, & \text{if } i \notin \{i_1, \ldots, i_k\} \\ 1, & \text{if } i \in \{i_1, \ldots, i_k\} \end{cases}.$$

Since there is a special covering for the set $S(f)$ under the special decomposition $d_n S(f)$), then according to Lemma 1.4 there is a special covering for the set $S(f)$ also under the special decomposition $(i_1, \ldots, i_k)I(d_n S(f))$.

It is easy to notice that the following ordered set
$$\{F_1^{\delta_1}, \ldots, F_i^{\delta_i}, \ldots, F_n^{\delta_n}\},$$
$$\text{for } \delta_i = \begin{cases} \alpha_i, & i \notin \{i_1, \ldots, i_k\} \\ 1 - \alpha_i, & i \in \{i_1, \ldots, i_k\} \end{cases},$$
will be a special covering for the set $S(f)$ under the special decomposition $(i_1, \ldots, i_k)I(d_n S(f))$.

In addition, obviously, $h(\delta_1, \ldots, \delta_n) = 1$. $\nabla$

To extend the operations of admissible changes, we added to them the operation of permuting components in an ordered pair of a special decomposition. New function is generated by the obtained special decomposition.

Let's find out how the clauses of a new function $h(x_1, \ldots, x_n)$ differ from the clauses of the given function $f(x_1, \ldots, x_n)$ when the components of some ordered pair, included in the special decomposition $d_n S(f)$, are permuted.

Suppose that for some $i \in \{1, \ldots, n\}$, the ordered pair $(F_i^0, F_i^1)$ under the special decomposition $d_n S(f)$, consists of following components:
$$F_i^0 = \{c_{l_1}, \ldots, c_{l_p}\} \text{ and } F_i^1 = \{c_{j_1}, \ldots, c_{j_q}\},$$
$$\text{where } \{c_{l_1}, \ldots, c_{l_p}\} \subseteq S(f) \text{ and } \{c_{j_1}, \ldots, c_{j_q}\} \subseteq S(f),$$

By the definition of these subsets, this means that:
- the literal $\bar{x}_i$ is included in all clauses of the set $c_{l_1}, \ldots, c_{l_p}\}$,
- the literal $x_i$ is included in all clauses of the set $\{c_{j_1}, \ldots, c_{j_q}\}$.
- the literals $\bar{x}_i$ and $x_i$ are not included in any other clauses.

As a result of permuting the components of the ordered pair $(F_i^0, F_i^1)$, the clauses of the set $\{c_{l_1}, \ldots, c_{l_p}\}$ move to the 1-domain of the corresponding decomposition, as well as the clauses of the set $\{c_{j_1}, \ldots, c_{j_q}\}$ move to the 0-domain of the corresponding decomposition.

We obtain the special decomposition $(i)I(d_n S(f))$, in which the $i$-th ordered pair has the form $(F_i^1, F_i^0)$. The remaining ordered pairs coincide with the corresponding ordered pairs of the decomposition $d_n S(f)$.
$$(i)I(d_n S(f)) = \{(F_1^0, F_1^1), \ldots, (F_i^1, F_i^0), \ldots, (F_n^0, F_n^1)\}.$$

Let's consider the procedure for generating the function $h(x_1, \ldots, x_n)$ in accordance with Section 2.4 based on the special decomposition $(i)I(d_n S(f))$.

Recall that we consider the set $S(f)$ as an ordered set.

The ordered set of clauses of the function $h(x_1, \ldots, x_n)$ will be denoted as
$$S(h) = \{c'_1, c'_2, \ldots, c'_m\}.$$

To form the $k$-th clauses $c'_k$, we search for subsets containing the element $c_k \in S(f)$ and form the set of literals, denoted by $l(c_k)$, as follows:

For any $j \in \{1, \ldots, n\}$ and $\alpha_j \in \{0,1\}$, if $c_k \in F_j^{\alpha_j}$, then the literal $x_j^{\alpha_j}$ is added to the formed set $l(c_k)$. Thus,
$$l(c_k) = \{x_j^{\alpha_j} / c_k \in F_j^{\alpha_j}, j \in \{1, \ldots, n\}, \alpha_j \in \{0,1\}\}.$$

In this case, the clauses of the function $f$ will be considered as elements of the set $S(f)$.

The clause $c'_k$ is formed as a clause composed of literals included the set $l(c_k)$.

Recall that $i$ is the number of the ordered pair of the decomposition $d_n S(f)$ whose components are permuted.

This means that after the permutation the subset $F_i^1 = \{c_{j_1}, \ldots, c_{j_q}\}$ is moved to the 0-domain, and $F_i^0 = \{c_{l_1}, \ldots, c_{l_p}\}$ is moved to the 1-domain of the resulting decomposition.

Since the function $h(x_1, \ldots, x_n)$ is generated by the resulting decomposition, then according to the procedure for forming clauses of this function, we will have the following:

- for any $c_{j_r} \in \{c_{j_1}, \ldots, c_{j_q}\}$, the new formed clause $c'_{j_r}$ of the function $h(x_1, \ldots, x_n)$ corresponding to $c_{j_r}$ will contain the literal $\bar{x}_i$.

- for any $c_{l_r} \in \{c_{l_1}, \ldots, c_{l_p}\}$, the new formed clause $c'_{l_r}$ of the function $h(x_1, \ldots, x_n)$ corresponding to $c_{l_r}$ will contain the literal $x_i$.

Let the ordered set
$$d_n S(h) = \{(H_1^0, H_1^1), \ldots, (H_i^0, H_i^1), \ldots, (H_n^0, H_n^1)\}$$
be the special decomposition of the set of new formed clauses. By formation procedure,

- the literal $\bar{x}_i$ is included in any of the clauses included in $H_i^0$,
- the literal $x_i$ is included in any of the clauses included in $H_i^1$.

At the same time, the clauses of the function $h(x_1, \ldots, x_n)$, not included in the subsets $H_i^0$ or $H_i^1$, coincide with the corresponding clauses of the function $f(x_1, \ldots, x_n)$.

Thus, we can state that if the following conditions are satisfied:

- $f(x_1, \ldots, x_n)$ is a Boolean function of $n$ variables represented in conjunctive normal form with the set of clauses $S(f) = \{c_1, c_2, \ldots, c_m\}$.
- the ordered set
$$d_n S(f) = \{(F_1^0, F_1^1), \ldots, (F_i^0, F_i^1), \ldots, (F_n^0, F_n^1)\}$$
is a special decomposition of the set $S(f)$.
- $d_n S(h)$ is a special decomposition obtained as a result of permutating the components of the ordered pair $(F_i^0, F_i^1)$ in the decomposition $d_n S(f)$.
- $h(x_1, \ldots, x_n)$ is a Boolean function generated by the special decomposition $d_n S(h)$.

Then, as a result of permutating the components of the ordered pair $(F_i^0, F_i^1)$, the following conditions will be satisfied.

- the clauses of the function $h(x_1, \ldots, x_n)$ not included in the subsets $H_i^0$ or $H_i^1$ coincides with the corresponding clauses of the function $f(x_1, \ldots, x_n)$.
- any clause included in the subset $H_i^0$ is obtained by replacing the literal $x_i$ in the corresponding clause of the function $f(x_1, \ldots, x_n)$ by the literal $\bar{x}_i$,
- in all clauses included in the subset $H_i^1$ is obtained by replacing the literal $\bar{x}_i$ in the corresponding clause of the function $f(x_1, \ldots, x_n)$ by the literal $x_i$.

In fact, we have shown that:

The function $h(x_1, \ldots, x_n)$ is obtained by replacing the literal $\bar{x}_i$ with the literal $x_i$ and the literal $x_i$ with the literal $\bar{x}_i$ in all clauses of the function $f(x_1, \ldots, x_n)$ that contain these literals.

Let's now study the properties of extended changes in a special decomposition.

Assume that we are given the special decomposition of a set $S$,
$$d_n S = \{(M_1^0, M_1^1), \ldots, (M_i^0, M_i^1), \ldots, (M_n^0, M_n^1)\}$$
such that the ordered set
$$c_n S = \{M_1^{\alpha_1}, \ldots, M_i^{\alpha_i}, \ldots, M_n^{\alpha_n}\}$$
is a special covering for the set $S$, where $\alpha_i \in \{0, 1\}$ for any $i \in \{1, \ldots, n\}$.

*Definition* 6.3. We say that the ordered set $d_n SG$ is generated by the decomposition $d_n S$ as a result of extended admissible changes, if the changes are performed in accordance with the following points:

- in addition to the admissible changes, a permutation procedure is also applied to some ordered pairs of the decomposition $d_n S$,

- if admissible changes are performed under some tuple of superscripts $(\sigma_1, \ldots, \sigma_n)$, and the permuting operation is applied to some $i$-th ordered pair of the decomposition under consideration, then admissible changes are continued under the tuple of superscripts

$$(\sigma_1, \ldots, 1\text{-}\sigma_i, \ldots, \sigma_n).$$

*Definition* 6.4. Let $f(x_1, \ldots, x_n)$ be a Boolean function of $n$ variables represented in conjunctive normal form with $m$ clauses, and there is an assignment tuple $(\sigma_1, \ldots, \sigma_n)$ such that

$$f(\sigma_1, \ldots, \sigma_n) = 1.$$

We will say that the function $h(x_1, \ldots, x_n)$ is generated by the function $f(x_1, \ldots, x_n)$ as a result of extended admissible changes, if:

- the special decomposition $d_n S(h)$ is generated as a result of extended admissible changes in the decomposition $d_n S(f)$,

- the function $h(x_1, \ldots, x_n)$ is generated by the special decomposition $d_n S(h)$.

*Theorem* 6.5. Let $f(x_1, \ldots, x_n)$ be a satisfiable Boolean function of $n$ variables represented in conjunctive normal form with $m$ clauses.

If the function $h(x_1, \ldots, x_n)$ is generated by the function $f(x_1, \ldots, x_n)$ as a result of extended admissible changes, then $h(x_1, \ldots, x_n)$ is a satisfiable function.

*Proof.* Let $d_n S(f)$ be a special decomposition generated by the function $f(x_1, \ldots, x_n)$. Since $f(x_1, \ldots, x_n)$ is a satisfiable function, then

$$f(\alpha_1, \ldots, \alpha_n) = 1$$

for some Boolean assignment tuple $(\alpha_1, \ldots, \alpha_n)$.

In addition, according to Theorem 2.3, the ordered set

$$c_n S(f) = \{F_1^{\alpha_1}, \ldots, F_i^{\alpha_i}, \ldots, F_n^{\alpha_n}\}$$

is a special covering for the set $S(f)$ under the decomposition $d_n S(f)$.

We will show that as a result of applying of any operation of the extended admissible change, we obtain a special decomposition such that there will be a special covering for the set $S(f)$ under this decomposition.

Let's consider two cases:

(a) the permuting operation is not applied during these changes.

In this case, according to Theorem 4.2, as a result of any operation we obtain a new special decomposition such that there is a special covering for the set $S(f)$ under this decomposition. In addition, the subsets included in the special covering have the same superscripts as the subsets of the original covering, $(\alpha_1, \ldots, \alpha_n)$.

This means that as a result $h(\alpha_1, \ldots, \alpha_n) = 1$ according to Theorem 2.3.

(b) if, during the extended admissible changes we need apply permuting procedure to some $i$-th ordered pair of the current decomposition, then according to definition we consider the tuple of superscripts of the subset included in special decomposition. Let it be $(\sigma_1, \ldots, \sigma_n)$.

After the permutation, obviously, $(\sigma_1, \ldots, 1-\sigma_i, \ldots, \sigma_n)$ will be the tuple of superscripts of the subsets in the special covering. According to Theorem 2.3, this means that the function generated by this special decomposition takes the value 1 if the variables are assigned the values
$$(\sigma_1, \ldots, 1-\sigma_i, \ldots, \sigma_n).$$

It is easy to see, that the final tuple of superscripts of the subset in the special covering will be a satisfying assignment for the function $h(x_1, \ldots, x_n)$. ∇

*Theorem* 6.6. Let $f(x_1, \ldots, x_n)$ and $h(x_1, \ldots, x_n)$ be arbitrary Boolean functions of $n$ variables represented in conjunctive normal form with $m$ clauses.

There are Boolean assignment tuples $(\sigma_1, \ldots, \sigma_n)$ and $(\delta_1, \ldots, \delta_n)$ such that
$$f(\sigma_1, \ldots, \sigma_n) = 1 \text{ and } h(\delta_1, \ldots, \delta_n) = 1.$$

Then, the function $h(x_1, \ldots, x_n)$ is generated by the function $f(x_1, \ldots, x_n)$ as a result of extended admissible changes, and also the function $f(x_1, \ldots, x_n)$ is generated by the function $h(x_1, \ldots, x_n)$ as a result of extended admissible changes.

*Proof*. Let the ordered set
$$d_n S(f) = \{(F_1^0, F_1^1), \ldots, (F_i^0, F_i^1), \ldots, (F_n^0, F_n^1)\}$$
be a special decomposition of the set $S(f)$, and let the ordered set
$$d_n S(h) = \{(H_1^0, H_1^1), \ldots, (H_i^0, H_i^1), \ldots, (H_n^0, H_n^1)\}$$
be a special decomposition of the set $S(h)$.

The functions $f(x_1, \ldots, x_n)$ and $h(x_1, \ldots, x_n)$ are satisfiable, hence the ordered set
$$c_n S(f) = \{F_1^{\sigma_1}, \ldots, F_i^{\sigma_i}, \ldots, F_n^{\sigma_n}\}$$
will be a special covering for the set $S(f)$ under the decomposition $d_n S(f)$, and the ordered set
$$c_n S(h) = \{H_1^{\delta_1}, \ldots, H_i^{\delta_i}, \ldots, H_n^{\delta_n}\}$$
will be a special covering for the set $S(h)$ under the decomposition $d_n S(h)$.

Let's proof that the function $h(x_1, \ldots, x_n)$ is generated by the function $f(x_1, \ldots, x_n)$ as a result of extended admissible changes. Consider the satisfiable tuples
$$(\sigma_1, \ldots, \sigma_n) \text{ and } (\delta_1, \ldots, \delta_n),$$
which also are the tuples of superscripts of the subsets included in $c_n S(f)$ and $c_n S(h)$, respectively.

We compare whether these tuples are the same.

- if $(\sigma_1, \ldots, \sigma_n)$ coincides with $(\delta_1, \ldots, \delta_n)$, then we use the procedure described in Theorem 4.3 to obtain the function $h(x_1, \ldots, x_n)$.

- for any $i \in \{1, \ldots, n\}$, if $\sigma_i \neq \delta_i$, then we assign the value $\delta_i$ to the element $\sigma_i$.

It is easy to see that this operation is equivalent to the permutation of the components of the ordered pair $(F_i^0, F_i^1)$, which is an admissible change. Therefore, we also permute the components of this ordered pair.

As a result of all these operations the special decomposition $d_n S(f)$ turns out to another special decomposition. In addition, there is a special covering under this decomposition such that the tuple of superscripts of the subsets included in it is $(\delta_1, \ldots, \delta_n)$.

On the other hand, the function generated by this special decomposition takes the value 1 if the variables are assigned the values $(\delta_1, \ldots, \delta_n)$.

Let's denote this function by $g(x_1, \ldots, x_n)$.

Since the function $g(x_1, \ldots, x_n)$ is generated by the function $f(x_1, \ldots, x_n)$, then it is enough to proof that the function $h(x_1, \ldots, x_n)$ is generated by the function $g(x_1, \ldots, x_n)$.

Thus, we obtained that $\delta_1, \ldots, \delta_n)$ is a satisfying assigning tuple for the functions
$$g(x_1, \ldots, x_n) \text{ and } h(x_1, \ldots, x_n):$$
$$f(\delta_1, \ldots, \delta_n) = 1 \text{ and } h(\delta_1, \ldots, \delta_n) = 1.$$

Obviously, the conditions of the Theorem 4.3 are satisfied for the functions $g(x_1, \ldots, x_n)$ and $h(x_1, \ldots, x_n)$.

This means the function $h(x_1, \ldots, x_n)$ is generated by the function $f(x_1, \ldots, x_n)$.

Therefore, the function $h(x_1, \ldots, x_n)$ is also generated by the function $g(x_1, \ldots, x_n)$. In a similar way we prove that the function $f(x_1, \ldots, x_n)$ is generated by the function $h(x_1, \ldots, x_n)$. ∇

*Theorem* 6.7. Let $f(x_1, \ldots, x_n)$ and $h(x_1, \ldots, x_n)$ be arbitrary Boolean functions of $n$ variables represented in conjunctive normal form with $m$ clauses.

There are assignment tuples $(\sigma_1, \ldots, \sigma_n)$ and $(\delta_1, \ldots, \delta_n)$ such that
$$f(\sigma_1, \ldots, \sigma_n) = 1 \text{ and } h(\delta_1, \ldots, \delta_n) = 1.$$

Then, the function $f(x_1, \ldots, x_n)$ generates the function $h(x_1, \ldots, x_n)$ as a result of extended admissible changes in no more than
$$c \times (n \times m)$$
elementary operations, for some constant $c$.

*Proof.* Suppose that the pair of (0,1)-matrices
$$((f)sM0, (f)sM1)$$
corresponds to the special decomposition $d_n S(f)$. According to Proposition 5.3, the pair of matrices can be generated by the function $f(x_1, \ldots, x_n)$ in no more than $c \times (n \times m)$ elementary operations for some constant $c$.

We will use the procedure described during the proof of Theorem 6.6.

So, let $g(x_1, \ldots, x_n)$ be a function which takes the value 1 if the variables are assigned the values $\delta_1, \ldots, \delta_n$. Recall that $g(x_1, \ldots, x_n)$ is the function generated by the ordered pair of matrices that is obtained as a result of permutation of some ordered pair of rows included in the ordered pair of (0,1)-matrices $((f)sM0, (f)sM1)$.

Obviously, the maximum number of elementary operations required for permuting the components of an ordered pair of rows included in $((f)sM0, (f)sM1)$ does not exceed the number $c \times m$ for some constant $c$.

Hence, maximum number of elementary operations required for permuting the components of all needed ordered pairs does not exceed the number $c \times (n \times m)$.

Since
$$h(\delta_1, \ldots, \delta_n) = 1 \text{ and } g(\delta_1, \ldots, \delta_n) = 1,$$
then according to Theorem 5.6, the function $g(x_1, \ldots, x_n)$ generates the function $h(x_1, \ldots, x_n)$ as a result of admissible changes under the assignment tuple $(\sigma_1, \ldots, \sigma_n)$ in no more than
$$c \times (n \times m)$$
elementary operations, for some constant $c$.

Combining these results, we can state:

Under the conditions of the theorem, the function $f(x_1, \ldots, x_n)$ generates the function $h(x_1, \ldots, x_n)$ as a result of extended admissible changes in no more than
$$c \times (n \times m)$$
elementary operations, for some constant $c$. ∇

Thus, using the concept of admissible changes we can implement the following:

- for any natural numbers $n$ and $m$, the set of satisfiable functions of $n$ variables, represented in conjunctive normal form with $m$ clauses, is partitioned into equivalence classes,

- the functions included in the same class have a common satisfiable assigning tuple.

- for any function included in a certain class, as a result of applying any admissible operation on this function, another satisfiable function included in the same class is obtained.

- any function of any equivalency class can be generated by an arbitrary function of the same class in polynomial time.

Extending the rules of the admissible changes,

For any natural numbers $n$ and $m$, all satisfiable functions of $n$ variables, represented in conjunctive normal form with $m$ clauses, are generated by each other in polynomial time.